\documentclass[aps,amsmath,amssymb,groupedaddress,preprint]{revtex4-1}
 \setcitestyle{numbers,square}

\usepackage{graphicx}% InclNanolett_Sachinude figure files
\usepackage{dcolumn}% Align table columns on decimal point
\usepackage{bm}
\usepackage[version=4]{mhchem}
\usepackage{amsmath,empheq}
\usepackage[utf8]{inputenc}
\usepackage[T1]{fontenc}
\usepackage{mathptmx}
\usepackage{etoolbox}
\usepackage{xcolor}
\usepackage{ulem}
\usepackage{siunitx}
\usepackage{braket}
\usepackage{tgbonum}
\usepackage{tcolorbox}

\DeclareSIUnit\elementarycharge{\text{\ensuremath{e}}}
\DeclareSIUnit\angstrom{\text {Å}}
\DeclareSIUnit\uc{\text {unit-cell}}

\usepackage{hyperref}% add hypertext capabilities
\hypersetup{
    colorlinks,%
    citecolor=blue,%
    linkcolor=blue,%
    urlcolor=blue
}

\begin{document}

%{\fontfamily{qcr}\selectfont}

%\preprint{}

\title{Orbital torque and efficient magnetization switching using ultrathin Co|Al light-metal interfaces: Experiments and modeling.}

\author{N. Sebe}
\author{A. Pezo}
\author{S. Krishnia}
%\email{skrishni@uni-mainz.de}
\thanks{Present address: Institute of Physics, Johannes Gutenberg University Mainz, Staudingerweg 7, 55128 Mainz, Germany}
\author{S. Collin}
\author{J.-M. George}
\author{A. Fert}
\author{V. Cros}
\email{vincent.cros@cnrs-thales.fr}
\author{H. Jaffrès}
\email{henri.jaffres@cnrs-thales.fr}
\affiliation{Laboratoire Albert Fert, CNRS, Thales, Universit\'e Paris-Saclay, 91767 Palaiseau, France}

\date{\today}

\begin{abstract}
{The emergence of the orbital degree of freedom in modern orbitronics offers a promising alternative to heavy metals for the efficient control of magnetization. In this context, identifying interfaces that exhibit orbital–momentum locking and an orbital Rashba–Edelstein response to an external electric field is of primary importance. In this work, we experimentally investigate the Co|Al system and extend the study to Co|Pt|Al structures. We show that inserting ultrathin Pt layers between Co and Al can significantly modify the orbital properties, highlighting the critical role of Co|Al orbital bonding in generating orbital polarization. We further model the orbital response of these systems using semi-phenomenological approaches and linear-response theory within the framework of density-functional theory.}
\end{abstract}

\maketitle

\section{Introduction \label{sec:intro}}

\subsection{General overview}

In the field of orbitronics, the orbital angular momentum (OAM), via the orbital Hall effect (OHE)~\cite{tanaka2008,Go2018intrinsic,go2020orbital,salemi2022first,lee2024,rang2025} and the orbital-Rashba Edelstein effect (OREE)~\cite{salemi2021quantitative,Johansson2021,Johansson2024theory}, have recently emerged as new observables able to tune quantum mechanical transport properties~\cite{cros2024}, enhance magnetic torques~\cite{krishnia2023large}, and boost functionalities for spintronic devices as spin-orbit torque MRAM (SOT-MRAM)~\cite{Lee2021,gupta2025harnessing,yao2025giant,Park2025}, THz spintronics emitters~\cite{Seifert2023,Xu2024,Liu2024,rxu2025,Tian2025,zou2025,Zhuang2025} or more recently in neuromorphic devices~\cite{gomes2025,Zhao2025}. That way, OHE and OREE appear perfect counterparts of the equivalent spin-based quantities, the respective spin-Hall (SHE)~\cite{sinova2015rmp} and Rashba effects (SREE)~\cite{sanchez2013spin,manchon2015perspectives} for controlling or switching the magnetization $\hat{\mathbf{M}}$ \textit{via} the action of the orbital torque or OT~\cite{Lee2021,Taniguchi2023,Yang2024} consisting of two components referred as to field-like (FL), $\hat{\sigma}\times \hat{\mathbf{M}}$ and damping-like (DL) symmetries $\hat{\mathbf{M}}\times\left(\hat{\sigma}\times \hat{\mathbf{M}}\right)$. Nevertheless, SHE and SREE strongly differing in that they rely on the use of heavy materials, as 5\textit{d} transition metals as Pt~\cite{guo2008intrinsic,Liu2011,Liu2012b,ralph2016}, W~\cite{Pai2012} or Ta~\cite{Liu2012}, because requiring a sizable spin-orbit coupling (SOC) interactions to promote the spin texture as a perturbative action as proposed with Schockley surface states~\cite{ishida2014,ishida2017,ishida2023} including \textit{sp}$^3$ basis for hexagonal stacking~\cite{ast2012}, limiting thus their amplitude for light elements. 

In that perspectives, the symmetry breaking and the lack of inversion and translation symmetries introduced at the interfaces between two light metals is fundamental for the generation of the orbital momentum locking (OML)~\cite{go2017toward} and the OREE conversion mechanism~\cite{Johansson2024theory,Nikolaev_2024_Nanoletters,pezo2025anatomy}. Electric-field induced orbital polarization was recently detected by electrical means in nonlocal devices (Cu/Al$_2$O$_3$ systems)~\cite{Gao2025}, even more recently by using surface acoustic wave or SAW (Ti|Ni)~\cite{Taniguchi2025};  or observed by more direct optical probe at the surface of light metallic layers of Ti~\cite{Choi2023} and Cr~\cite{Lyalin2023} and by local electronic probe on Ti~\cite{idrobo2024}. 

Already at the origin of many important fundamentals aspects in condensed matter physics as surface and perpendicular magnetic anisotropy (PMA),  Dzyaloshinskii–Moriya interactions (DMI), the role of the orbital angular polarization in OREE relies on the OML at Fermi surfaces~\cite{go2017toward}, that is a peculiar chiral texture in the Brillouin zone wherein the spin quantum direction is fixed along the direction transverse to the electron wavevector. Under the action of an in-plane electric field, such specific chiral texture enables the generation of a non-equilibrium orbital polarization density, or orbital accumulation~\cite{manchon2015perspectives,Johansson2024theory} enable to interact with the magnetization of a thin ferromagnetic layer, thereby enhancing the magnetic torques, named the orbital torque (OT). The latter may exceed the efficiency that of the spin-Hall effects (SHE) of heavy metals~\cite{gupta2025harnessing,yao2025giant}. Another relevant example is also played by quantum materials, like topological insulators, which can exhibit equivalent orbital properties at their interfaces~\cite{park2012chiral,Xie2014} and able to provide significant orbital responses~\cite{pezo2024theory,Lee2025}.

While spin momentum locking (SML) and the resulting spin-REE~\cite{Sanchez2013} typically requires strong SOC to emerge, OML is anticipated to be less restrictive as it is not directly linked to SOC~\cite{go2017toward,ding2022observation, Go_2021_ORE}, at least for the orbital generation process. This leads to new fundamental paradigms described by recent models proposed for bulk materials ~\cite{salemi2021quantitative,Sala_2022_3d5d4f,kawakami2024,Vignale2025,culcer2025,Wang2025,Jin2025,Bonythese2025} or interfaces~\cite{lyalin2025,ning2024} where the orbital polarization, whatever its origin, is coupled to the spin by SOC. More recent theories have invoked charge vorticity~\cite{okano2019} and/or gradient effects to assert for a local origin of the electric-field generated orbital polarization~\cite{valet2025quantum}, as further observed experimentally~\cite{Nozaki2025,yi2025}. However, when it comes to the understanding the physical microscopic origin of the orbital torque (OT), it remains challenging to experimentally differentiate between the proposed models. This is because spin and Orbital effects indeed share the same symmetries~\cite{Go_2020_OTgen}. Nevertheless, two main differences between spin and orbital accumulations can be made. First, the amplitude of the OT requires a small SOC inside the thin ferromagnetic layer, meaning changing the ferromagnetic material modifies the strength of the torques~\cite{Kim_2021_nontrivial}. This may result in specific properties where the interaction between the orbital polarization and the exchange in the ferromagnet can extend over large distances~\cite{Hayashi2023,Liu2023,go2023long,ding2022observation,bony2025quantitative} through hot-spot channels. Second, transmission of orbital accumulation is more sensitive on the crystallinity of the interfaces than transmission of spin accumulation~\cite{Go_2020_OTgen}.

In this study, we present various trends in electrical transport measured in different series of samples comprising a Co|Al interfaces as previously experimentally demonstrated~\cite{krishnia2023large}. We will point out how our measurements lead to the conclusion that an OREE exerting a strong spin-orbit torque (SOT) onto the Co magnetization. The article is organized as follows: Section II provides a general introduction on the orbital contribution to the SOT and OT, highlighting key concepts and challenges related to the torque experiments. Section III describes sample properties covering structural, PMA, spin-dependent transport and anomalous Hall effects (AHE). Section IV presents SOT and OT measurements on the different samples series using second-harmonic techniques as well a semi-classical modeling of SHE spin-injection. Section V focuses on the electronic band structure calculations obtained and OREE linear response via density functional theory (DFT). Finally, section VI discusses magnetization switching experiments highlighting how the contribution from current-induced orbital polarization helps for the threshold current density reduction.

\subsection{Description of the orbital torque at FM/LM interfaces~\label{sec:SOT}}

In this first section, we introduce the key concepts and prerequisites for benchmarking the orbital torques (OT) generated by the OREE at a ferromagnet/light metal (FM/LM) interface. These are based on our recent experimental data already published ~\cite{krishnia2023large} as well as the results presented in the following sections. An important consideration is making a fair comparison between the efficiencies of SHE and OREE, as these two quantities cannot be compared due to their differences in units. Indeed, differences between the mechanisms of SHE vs. REE are multiple including the fact that REE primarily arises from an angular momentum polarization densities or \textit{accumulation}, while SHE and OHE are associated to the flow of these quantities. Furthermore, the distinction between REE and SHE is mainly linked to the \textit{intraband} vs. \textit{interband} part of the response tensor, which results respectively in a field-like (FLT) or  damping-like (DLT) torque component. 

Evaluating and separating OT and SOT is complex task. Nevertheless, any orbital system is generally less sensitive to the exchange-correlation terms~\cite{Nikolaev_2024_Nanoletters}, enables us to draw particular conclusions: 3\textit{d} FM|LM interfaces with reduced SOC appear as a toy model to investigate the fundamentals of the OT; unlike FM|heavy metals systems, wherein SOC competes with exchange in strength, thus strongly impacting the both the torques~\cite{go2020theory} and the dynamics of the magnetization reversal~\cite{devda2025}.

Another important aspect to be considered is \textit{(i)} the role of source terms originating from the flow of a local \textit{out-of-equilibrium} (spin or orbital) polarization, $\hat{\mu}_{L,S}(z)$ and \textit{(ii)} their spatial-temporal dynamics in the FM layers from to the exchange-torque effects. Such description cannot be handled correctly by first-principle calculation techniques as it involves coupled diffusion equations, which are typically treated either using coupled semi-classical (\textcolor{blue}{Suppl. Info. III}) or by atomistic equations~\cite{devda2025}. Furthermore, as recently proposed~\cite{Valet1,valet2025quantum}, a thorough and consistent understanding of the OT  requires a deep analysis of the local current-in-plane (CIP) gradient because linked to \textit{interband} transitions, source of angular momentum polarization.

From the viewpoint of the generalized Landau-Lifschitz-Gilbert equation, describing the magnetization dynamics, the SOT vector $\boldsymbol{\hat{\Gamma}}_{SOT}$ defines an effective effective magnetic field $\bf{B}^{eff}_{SOT}$ acting on the local magnetization $\mathbf{M}(z)$ such that:
\begin{equation}
    \boldsymbol{\Gamma}_{SOT}(z)=\gamma~ \mathbf{B}^{eff}_{SOT} \times \mathbf{M}(z)
\end{equation}
with $\gamma = \left(\frac{e}{2 m*}\right)$ the gyromagnetic factor ($m^*$ is the effective mass) where the 'hat' symbol defines a 2-dimensional (2D) vector normal to $\mathbf{M}$ in the angular-momentum space, defining thus both a field-like and damping-like SOT and OT components. In the steady-state of magnetization dynamics, \textit{i.e.} in a timescale well larger than the typical electronic relaxation times, the \textit{local torque} writes $\boldsymbol{\hat{\Gamma}}_{SOT}(z)=\frac{\partial \mathbf{M}(z)}{\partial t}  =-\boldsymbol{\hat{\Gamma}}_s(z)$ with $\boldsymbol{\hat{\Gamma}}_s(z)= \frac{\partial \mathbf{\hat{\mu}}_s(z)}{\partial t}=-\hat{Q}(z)$, opposite to the torque acting on the generated spin accumulation as a reactive effect. It is directly linked to \textit{(i)} the local spin accumulation $\hat{\mu}_s(z)$ and to \textit{(ii)} the spin-current influx, $\hat{Q}(z)=-\nabla \hat{J}_{\sigma}(z)$ with the assumption of a small SOC strength compared to the 3\textit{d} exchange-correlation interaction $\Delta_{xc}=\mathbf{B}_{xc}.\mathbf{M}$.
The validity of the quantum correspondence has been given in Ref.~\cite{go2020theory}. It turns out that, under the assumption of small SOC, $\mathbf{B}^{eff}_{SOT} =\left(\frac{\Delta_{exc}}{M_s t_F}\right)\check{\hat{\mu}}_s$ considering the 2D-spatial spin accumulation $\check{\hat{\mu}}_s$ (the 'checkmark' $\checkmark$ symbol has been added to note the 2D-spatial character in unit of m$^{-2}$ - the double 'checkmark' and 'hat' symbol describes two dimensionality for the transverse angular momentum component and two-dimensionality in space at the Co|Al interface). 

\vspace{0.1in}

Electronic band structure calculations of Co|Al interfaces~\cite{Nikolaev_2024_Nanoletters,pezo2025anatomy} revealed a much stronger in-plane OAM $\hat{L}$ at equilibrium together with an out-of-equilibrium 2D orbital polarization $\check{\hat{\mu}}_{L}= \chi_{yx} E_x$ (in unit of $\hbar/m^{2}$) compared to their spin counterparts. This suggests that a conversion process between OAM $\hat{L}$ and spin $\hat{\sigma}$ inside the ferromagnet is essential for the observation of a significant OT. Such interconversion may be considered as a perturbation to the first order occurring during the torque process itself~\cite{Nikolaev_2024_Nanoletters}. In the light of this note, one can approximate $\check{\hat{\mu}}_s$ to $\check{\hat{\mu}}_s\simeq \left(\frac{\Delta_{SOC}}{\Delta_{xc}}\right) \check{\hat{\mu}}_L$ to give \textit{in fine} the \textit{orbital torque} field:

\begin{equation}
    \mathbf{B}^{eff}_{OT} =\left( \frac{\Delta_{SOC}}{\hbar M_s t_F}\right) \check{\hat{\mu}}_L
    \label{equ:equivalentSOTfield}
\end{equation}
with $\Delta_{SOC}=70~$meV the characteristic 3\textit{d} SOC strength for Co. However, based on the form of Eq.~[\ref{equ:equivalentSOTfield}], it is not possible to clearly distinguish between the exchange-torque of an orbital origin, evaluated by a perturbation calculation at the lowest order, and an additional torque contribution from the SOC-field itself.

\vspace{0.1in}

How then can we compare the respective efficiencies of SHE and OREE torques ($\xi_{DL/FL}$)? Even though SHE is more associated with the DL torque due to its \textit{dynamic} nature and OREE primarily contributes to the FL component, such a comparison is required to benchmark our experimental data against the linear response from density functional theory (DFT). Our answer is based on the following arguments:

\textit{(i)} the SHE current from Pt scales as $J_{SHE}=\left(\frac{\hbar}{2e}\right)\sigma_{SHE} \mathbf{E}$ with $\sigma_{SHE}$ the (intrinsic) spin Hall conductivity (SHC) and $\mathbf{E}$ the electric field.

\textit{(ii)} Correspondingly, concerning OREE~\cite{Fert2019} (as for the spin-REE), the equivalent OREE current is given by $J_{OREE}=\left(\frac{\hbar\hat{\hat{\mu}}_s}{\tau_{F}}\right)=\gamma_F \check{\hat{\mu}}_s$. $\tau_{F}=\left(\frac{\hbar}{2 \gamma_{F}}\right)$ is the typical carrier spin lifetime - or escape time - onto the localized Rashba state~\cite{Sanchez2013} at the Co|Al interface with $\gamma_{F}$ is the typical broadening energy at the Fermi level.

\textit{(iii)} From the perturbation calculation to the lower order, the 2D spin density $\check{\hat{\mu}}_s$ is related to the generated orbital density $\check{\hat{\mu}}_L=\chi_{xy}^L \mathbf{E}$ and estimated from the typical ratio $\epsilon_{SO}=\left(\frac{\Delta_{SOC}}{\Delta E_{CF}}\right)\simeq \left(\frac{\Delta_{SOC}}{\Delta_{xc}}\right)\simeq 0.1$ between the SOC strength over the crystal-field (or exchange) splitting $\Delta E_{CF}\simeq \Delta_{xc}\simeq 1~$eV/atom. $\chi^L$ is called the OREE tensor~\cite{Johansson2021}.

Equating $J_{SHE}$ and $J_{OREE}$ gives equivalent OREE and SHE torque magnitude. This results in:

\begin{eqnarray}
    \chi_{xy}^L\simeq  \left(\frac{\hbar \sigma_{SHE}}{2e\epsilon_{SO}\gamma_F}\right)\simeq 10^{11} (\hbar)/(V.m) \nonumber \simeq \\
    \simeq 10^{-9} (\hbar)/(\text{atom}~V m^{-1})
    \label{chi_ordermagnitude}
\end{eqnarray}
considering that $\sigma_{SHE}\simeq 10^5$~S/m for Pt~\cite{rojas2014spin} and $\gamma_F \simeq 0.1$~eV corresponding to a relative short escape time at a metal interface~\cite{Sanchez2013} (\textit{i.e.} spin or escape time of the order of 10~fs onto the Rashba states in agreement with the Co resistivity). 

\vspace{0.1in}

The out-of-equilibrium orbital polarization $\check{\hat{\mu}}_L$ emerging from the OML may be evaluated using Kubo's formalism as a sum of two contributions from the respective \textit{intraband} and \textit{interband} components respectively; although the latter, resulting from the virtual coupling between different bands, is generally expected much smaller (vanishing in the limit of time reversal symmetry or TRS). A value of $\mathbf{E}$ of the order of $2.5\times10^4$~V/m, corresponding to a current density in Pt of  $10^{11}$~A/m$^2$ (with a Pt resistivity equal to $\rho_{Pt}=25\mu\Omega.$cm) yields an orbital moment density $\check{\hat{\mu}}_L$ approaching $2.5\times10^{-5}(\hbar)$/atom resulting in an equivalent FLT approaching $\mathbf{B}_{FL}\simeq3$~mT, for $\left(Ms~t_F\right)\simeq 1\mu_B/$atom, a value in very good agreement with the experimental value~\cite{krishnia2023large}. 

\vspace{0.1in}

\textit{(iv)} The use of Density functional theory allows a more fair quantitative evaluation of the so-called 'torquance' $t_{\alpha\beta}$~\cite{freimuth2014}, will be evaluated from the first-principle techniques within the linear response theory framework:
\begin{equation}  \hat{\Gamma}_{SOT,\alpha}=t_{\beta\alpha}~E_\beta
\end{equation}
For an electric field $E_x$ applied along $\hat{x}$ and the magnetization $M_z$ along $z$, $t_{\beta\alpha}$ can be separated into its respective longitudinal (FL, $t_{xx}$) and transverse (DL, $t_{xy}$) components. However, we cannot differentiate between the respective spin and orbital contributions without further evidence of the OML vs. SML signatures in the electronic band structure.

\section{Results}

\subsection{Sample growth and characterization}

\textit{Structural and magnetic properties}

\vspace{0.1in}

In this section, we investigate the impact of inserting atomic-size Pt layer at Co|Al interface and compare the results to our recent experimental results in Pt|Co$\parallel$Al|Pt and Pt|Co|AlOx series. Multilayers made from two different series, Ta5|Pt8|Co($t_{Co}$)|Al(1.4-3)|Pt3 (series A) and Ta5|Pt8|Co(0.9)|Pt($t_{Pt}$)|Al(3)|Pt3 (series B) were then grown by sputtering on Si/SiO$_2$(180~nm) insulating substrates at the base pressure of $5\times 10^{-8}$~mbar. Two additional samples Ta5|Pt8|Co(0.9)|Al(1)Ox (sample C) and Ta5|Pt8|Co(0.9)|Al(4)Ox (sample D) with top oxidized Al capping were grown to benchmark the measured torques. C and D can be considered as two references capped by AlOx instead of Pt. Note that, in sample D, the Al oxidation is not total with the result that the Co|Al interface is made with metallic Al.

\vspace{0.1in}

In our recent publication~\cite{krishnia2025}, we employed complementary experimental techniques including X-ray reflectivity (XRR), X-ray photoelectron spectroscopy (XPS), Scanning Transmission Electron Microscopy (STEM) using energy dispersive X-ray spectroscopy (EDX). These experiments clearly demonstrate the high quality of the Co|Al|Pt and Co|AlOx multilayers, characterized by flat interfaces and very limited interdiffusion. These  findings were complemented by magnetic characterizations using SQUID magnetometry and element-selective x-ray circular dichrosim (XMCD) which revealed that the dependence of the saturation magnetization on the Al-thickness ($t_{Al}$)  for the sub-nanometric Co layer, exhibited a gradual drop of $M_s$ as $t_{Al}$ exceeded $0.7~$nm corresponding to its in-depth self-saturation oxidation threshold.~\cite{krishnia2025}. 

We assigned this behavior to three different factors supported by DFT calculations: \textit{(i)} the reduction of the spin moment of Co at the vicinity of the metallic Al, and \textit{(ii)} the increase of the atomic cell volume for 'thick' Al accompanying an in-plane tensile strain. The ensemble of these features explain that the magnetic moment of 0.9~nm ultrathin Co in Pt(8)/Co(0.9)/Al(3) is about only 950 emu/cm$^3$ which appears to be reduced compared to the Co bulk value (1400 emu/cm$^3$). In addition to this, we also found a double sign change of the magnetic anisotropy when $t_{Al}$ is varied from 0.7 to 1.2~nm, evidencing a double magnetic reorientation from PMA to in-plane magnetized and recovering PMA for $t_{AL}>1.2$~nm~\cite{krishnia2025}.

In the next sections, we address the magnetic and transport properties of samples for both series A (Pt|Co|Al(t)|Pt) and series B (Pt|Co|Pt(t)|Al|Pt involving the interleaved atomic Pt layers. Thanks to the high quality of the bottom Pt|Co and the top Co|$\tilde{\text{Pt}}$|Al interfaces, all theses samples exhibit strong PMA properties. Saturation magnetization at room temperature for $t_{\tilde{\text{Pt}}}=0.25, 0.5, 0.75, 1, 2, 3~$nm is respectively 1035, 1060, 1120, 1120, 1390 and 1440 emu/cm$^3$ (\textcolor{blue}{see table I in the Suppl. Info. I}) is representative of the increase of the magnetic moments due to magnetic proximity effects in $\tilde{\text{Pt}}$ in Pt(8)|Co(0.9)|$\tilde{\text{Pt}}$(t$_{Pt}$)|Al(3) (B series).

\vspace{0.1in}

\textit{Spin-polarized transport (AHE)}

\vspace{0.1in}

Before investigating transverse spin-currents as source of SOT and OT, we first address the spin-dependent transport properties in the current in-plane (CIP) geometry. Analyses of CIP data rely on the Anomalous Hall Effect (AHE) \textit{via} the measurement of the transverse resistance $R_{xy}=R_{AHE}$ \textit{vs.} $M_z$. Indeed, $R_{AHE}$ appears to be a powerful probe for determining the profiles of both charge and spin currents, along with their spatial gradients, which are strongly influenced by the scattering properties in the vicinity of the interfaces. In an extended spin-dependent picture~\cite{camley1,camley2,Butler1995,THDang2020anomalous,bony2025quantitative} of the Fuchs-Sondheimer approach~\cite{Fuchs1938,Sondheimer1952}, these overall properties are scaled by different parameters as the local spin-dependent conductivity ($\sigma_{xx}^{\uparrow \downarrow}$)~\cite{Butler1995}, linked to the electron mean-free path ($\lambda^{\uparrow \downarrow}$), the electronic spin-dependent transmission ($T^{\uparrow\downarrow}$), the reflection ($R^{\uparrow\downarrow}$), spin-loss ($\delta$) and the interface specularity ($\textit{sp}$)~\cite{stewart2003interfacial}. Furthermore, most of the parameters mentioned above, except the interfacial electronic specularity, enters in the fundamental principles of SOT/OT CPP currents. As a result, the extracted physical quantities are used to fit the SOT|OT experimental data. (see \textcolor{blue}{Suppl. Info. III}). 

\begin{figure}[h!]
    \centering
\includegraphics[width=0.85\textwidth]{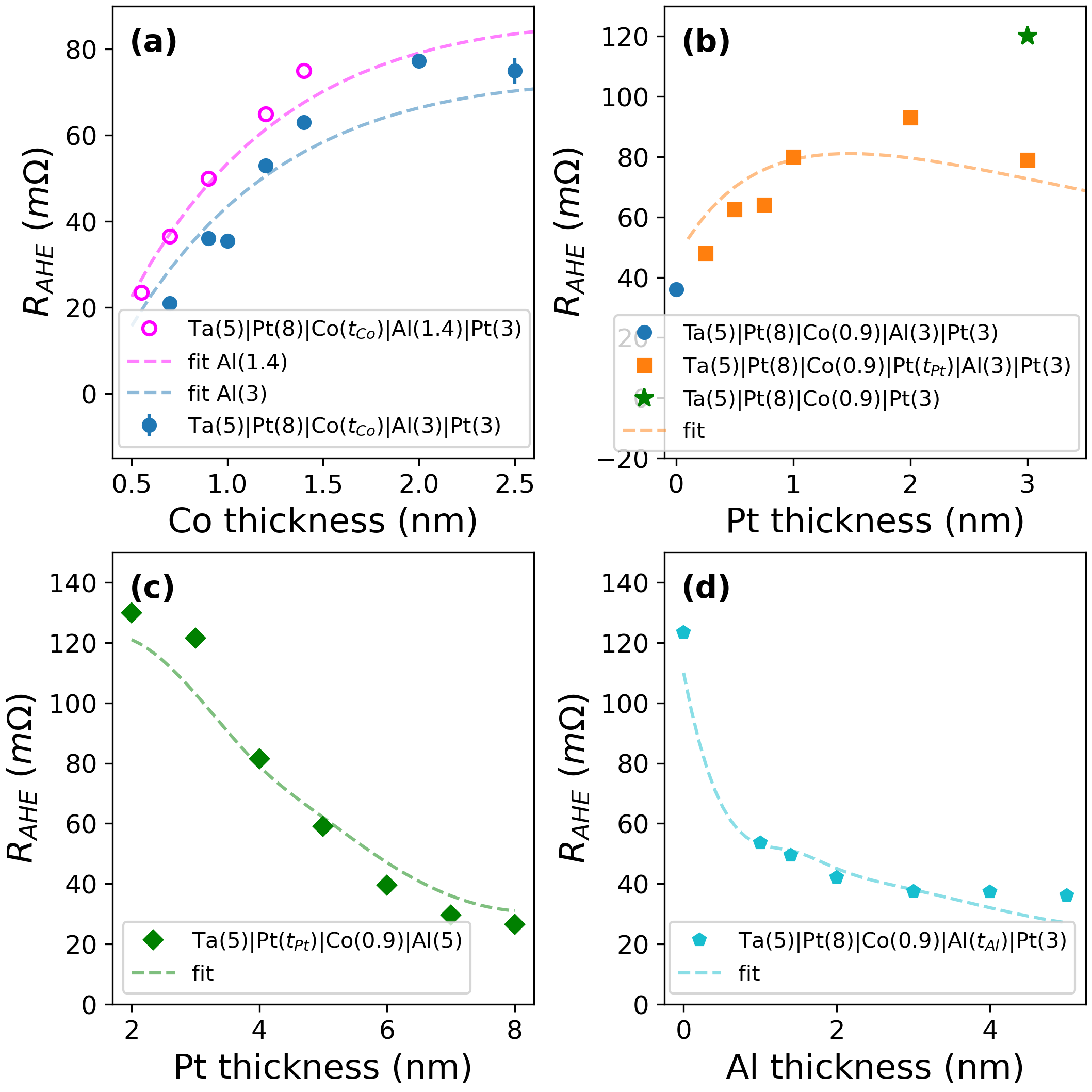}%
    \caption{\label{fig:AHE} AHE resistances measured in series of multilayers. (a) Two series of samples were measured varying Co thickness in $\mathrm{Ta(5)|Pt(8)|Co(}t_{Co}\mathrm{)|Al(1.4 \text{or} 3)|Pt(3)}$. We then measured AHE (b) in samples with Pt inserted between Co and Al ($\mathrm{Ta(5)|Pt(8)|Co(0.9)|Pt(}t_{Pt}\mathrm{|Al(3)|Pt(3)}$). The thickness of bottom Pt layer is varied in $\mathrm{Ta(5)|Pt(}t_{Pt}\mathrm{)|Co(0.9)|Al(5)}$ (c).(d) AHE resistance as a function of Al thickness in $\mathrm{Ta(5)|Pt(8)|Co(0.9)|Al(}t_{Al}\mathrm{)|Pt(3)}$.}
\end{figure}

\vspace{0.1in}

In more details, the AHE resistance writes $R_{AHE}=\frac{\int \sigma_{xy}dz}{\left(\int \sigma_{xx}dz\right)^2} \simeq \frac{\sum_l \sigma_{xy}^{(l)}~t^{(l)}}{\left(\sum_l \sigma_{xx}^{(l)}~t^{(l)}\right)^2}$ with $(l)$ the index of the layer after having possibly used a discretization procedure with $t^{(l)}$, $\sigma_{xx}^{(l)}$ and $\sigma_{xy}^{(l)}$ being respectively the thickness, the longitudinal conductivity and the Hall conductivity of the corresponding layer (l) within the structure. In this scheme, any additional spin-orbit assisted scattering of a spin-polarized current in a certain layer (l) by \textit{spin-injection proximity effect}, as played by the Pt HM, leads to the generation of an additional Hall currents at the length scale of its mean-free path. 

In Fig.~\ref{fig:AHE}, we show the $R_{AHE}$ acquired on different sample series \textit{vs.} layer thicknesses. Fig.~\ref{fig:AHE}(a) displays the dependence of AHE on the Co thickness ($t_{Co}$) in $\mathrm{Ta(5)|Pt(8)|Co(}t_{Co}\mathrm{)|Al(1.4)|Pt(3)}$ and $\mathrm{Ta(5)|Pt(8)|Co(}t_{Co}\mathrm{)|Al(3)|Pt(3)}$ emphasizing an increases of the AHE vs. $t_{Co}$ for the two main series (Al 1.4 and Al 3~nm). Such $R_{xy}$ behavior manifests first \textit{(i)} the gradual promotion of the spin-current polarization $P_s^{Co}$ for the ultrathin Co~\cite{bony2025quantitative}. Indeed, at low Co thickness, $P_s^{Co}$ is strongly reduced because of a loss of specularity in a Fuchs-Sondheimer scheme of interfacial diffusion processes, also possibly involving spin-flips, and secondly \textit{(ii)} the saturation of $P_s^{Co}$ for thick Co ($t_{Co}>2.5$~nm) limited to its value in the bulk ($\beta^{Co}\simeq 0.65\pm 0.05$) leads to a decrease of $R_{xy}$ with the expected $1/t_{Co}$ law. We note that the AHE appears always larger for 1.4~nm Al compared to 3~nm Al. Our data and simulations show that this feature has not to be assigned to a strong additional current shunt in thin Al. The origin is a quenching, through the interleaved Al layer, of the electronic transmission from Co to Pt region of enhanced spin-orbit assisted skew-scattering processes~\cite{dang2020anomalous,bony2025quantitative}. Alongside, Fig.~\ref{fig:AHE}(d) displays the AHE resistance $R_{xy}$ for $\mathrm{Ta(5)|Pt(8)|Co(0.9)|Al(}t_{Al}\mathrm{)|Pt(3)}$ series. One notices a first rapid drop of $R_{xy}$ from 120~m$\Omega$ to 55~m$\Omega$ when a thin 1.4~nm Al is inserted at the top 'Co|Pt' interface. Such behavior is also assigned to a decrease of the electron spin-transmission between Co and Pt through Al.

Fig.~\ref{fig:AHE}(b) corresponds to the AHE for the $\tilde{\text{Pt}}$ interleaved series (series B). When $\tilde{\text{Pt}}$ is gradually inserted between Co and Al, $\mathrm{R_{AHE}}$ gradually enhances from 36~m$\Omega$ (Pt|Co|Al) to 
48~m$\Omega$ for 0.25~nm $\tilde{\text{Pt}}$ and to 93~m$\Omega$ for 2~nm $\tilde{\text{Pt}}$ where AHE reaches its maximum before decreasing due to current shunting effects in $\tilde{\text{Pt}}$. Such AHE enhancement in intercalating $\tilde{\text{Pt}}$ originates the equivalent enhanced skew-scattering processes experienced by the spin-current $P_s^{Co}$ penetrating $\tilde{\text{Pt}}$. This process is very efficient as the 'bare' spin-Hall angle of Pt, here measured by AHE, exceeds $\theta_{SHE}^{Pt}=0.22$ ($\sigma_{SHC}^{Pt}\simeq 8.5\times 10^3$~S/cm ($\sigma_{SHC}^{Pt}\simeq 3.4\times 10^3$~S/cm considering the spin-memory loss $\delta=0.4$~\cite{rojas2014spin}) much more than the one for Co (0.02) leading to AHE angle of the order of 0.007 for Co~\cite{bony2025quantitative}.

The rise of AHE from Co|Pt interfaces becomes even more apparent in $\mathrm{Ta(5)|Pt(8)|Co(0.9)|Pt(3)}$ free of Al ($\mathrm{R_{AHE}}$= 120~m$\Omega$) because avoiding the additional shunt from the top capping layer in the series B. Regarding this series, the maximum value in the $\mathrm{R_{AHE}}$ observed at the vicinity of $t_{Pt}=2$~nm is a consequence of a finite in-plane spin-transport in Pt occurring on a typical lengthscale close to the electronic the mean-free path, location of the enhanced skew-scatterings. Fig.~\ref{fig:AHE}c) shows how AHE is dependent on the thickness ($t_{Pt}$) of bottom Pt SHE layer. AHE decreases from 130~m$\Omega$ to 26.5~m$\Omega$ when the bottom Pt thickness is increased from 2 to 8~nm. The thicknesses of Pt in those samples are typically larger than the spin mean-free path ($\lambda^{Pt}\simeq$1.5~nm) extracted from the analysis of Fig.~\ref{fig:AHE}b). It results in a decrease of AHE vs. $t_{Pt}$ due to the current shunt in Pt.

\vspace{0.1in}

The ensemble of the physical parameters used for the fitting procedure, among which the respective conductivity of Co ($\sigma_{xx,Co}^{\uparrow}=7.4\times10^6$~S/m and $\sigma_{xx,Co}^{\downarrow}=1.6\times10^6$~S/m corresponding to respective mean-free path $\lambda_{Co}^{\uparrow\downarrow}\simeq$~ 7 and 2 ~nm) and Pt ($\sigma_{xx,Pt}=5\times10^6$~S/m, $\lambda_{Pt}\simeq$~2~nm), presented in Figs.~\ref{fig:AHE} are given in the table given in the \textcolor{blue}{Suppl. Info. II}.

\subsection{sOT and OT from second harmonic methods \label{f-2f}}

\textcolor{red}{In this section, we} investigate the characteristics of SOT and OT in the different series. SOT magnitude was acquired using standard harmonic-Hall measurements techniques. For the determination of the respective damping-like ($H_{DL}$) and field-like ($H_{FL}$) torque components, a magnetic field is swept either in the direction parallel to the current injected (DL) or in the plane transverse to it (FL) along an axis disoriented of 4° out of the plane of the Hall bar in order to keep the structure the magnetization in a monodomain state. These two experimental configurations are respectively called DL and FL geometries when $R_{AHE}\gg R_{PHE}$. The shape of the second harmonic signal $V^{2\omega}$ is fitted on the whole field range to extract the amplitude of the torques. 

The respective FL and DL efficiencies of the integrated torque in the volume write:

\begin{equation}
\xi_{DL/FL} = \left(\frac{2e}{\hbar}\right)\left(\frac{\mu_0 H_{DL/FL}M_s t_{Co}}{J_{Pt}}\right)
\end{equation}
where $t_{Co}$ and $M_s$ are the thickness and saturation magnetization of Co previously determined.

\vspace{0.1in}

\textit{Co|Al interfaces: case of the pure OT}

\vspace{0.1in}

\begin{figure}
\includegraphics[width=0.85\textwidth]{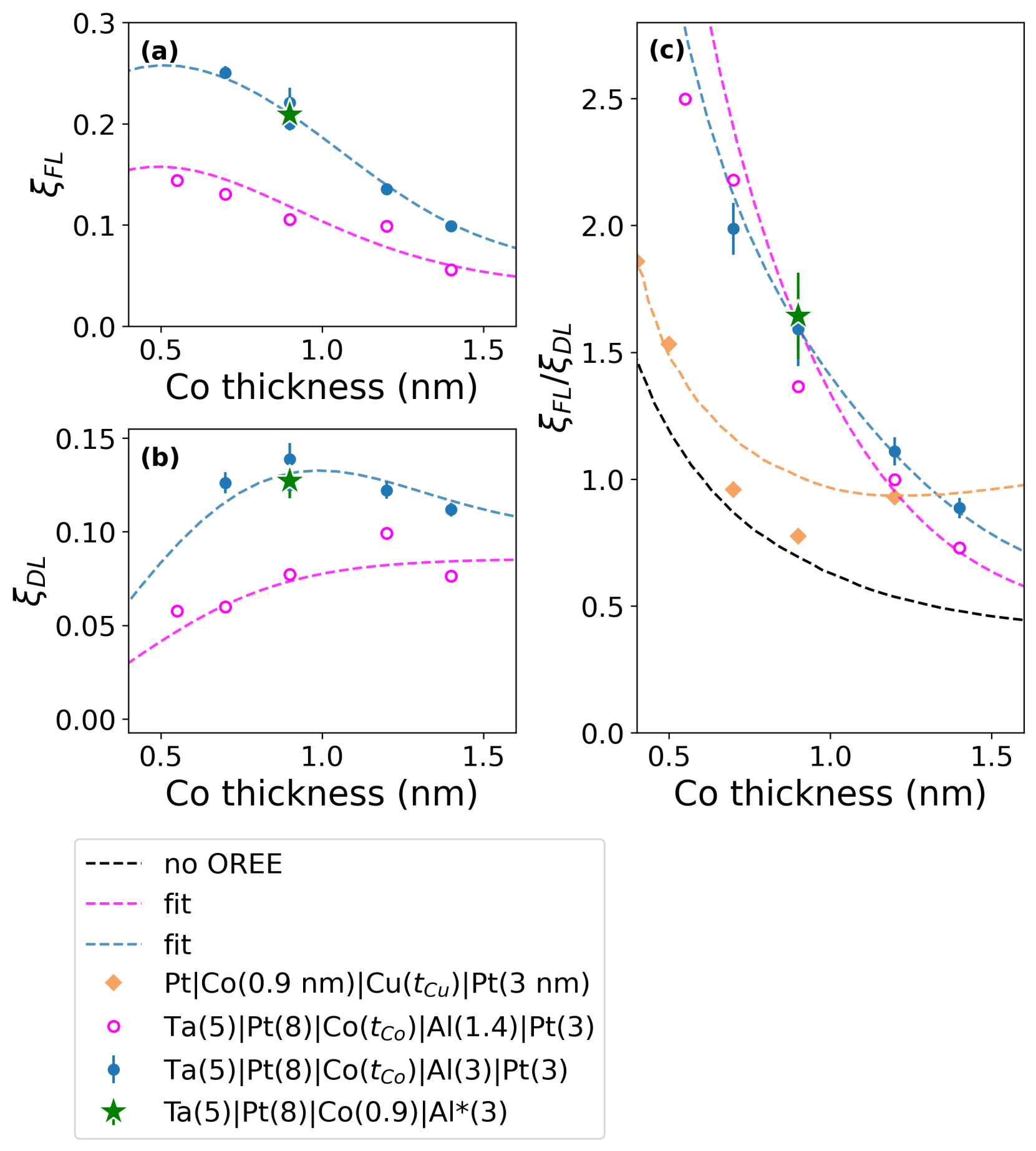}%
\caption{Co thickness dependence of the damping like torque efficiency \textbf{(a)}, the field like torque efficiency \textbf{(b)} and the ratio between the two torques \textbf{(c)}. For each of the plots, we compare two series of sample with different Al thickness: Ta(5)|Pt(8)|Co($t_{Co}$)|Al(1.4-3)|Pt(3). The orange plot represents the reference sample replacing Al by Cu. The green star represents the sample Ta(5)|Pt(8)|Co($t_{Co}$)|Al(4)Ox free of the top Pt. The black dot line represent the line without any Rashba contribution. Data from Al(1.4) series are adapted from Ref. ~\cite{krishnia2023large}.}
\label{fig:Efficiency} 
\end{figure}

We first discuss the results of the series A free of Pt interleaved layer at Co|Al interfaces. Fig.~\ref{fig:Efficiency} exhibits the torque efficiencies in $\mathrm{Ta(5)|Pt(8)|Co(}t_{Co}\mathrm{)|Al(1.4\ or\ 3)|Pt(3)}$ series when $t_{Co}$ is varied from 0.5 to 1.5~nm. Data from 1.4~nm Al were already published in a previous reference~\cite{krishnia2023large} and we refer to it to describe our major results obtained for 3~nm Al. Fig.~\ref{fig:Efficiency}(a) display the strong increase of field-like component ($\xi_{FL}$) by almost a factor of two when the Al thickness is increased from of 1.4~nm to 3~nm. For $t_{Co}\simeq 1.5$~nm (what we call thick Co), $\xi_{FL}$ is about 0.05 in the range of values expected. On decreasing Co thickness down to 0.5~nm still in the ferromagnetic domain, $\xi_{FL}$ reaches respectively 0.15 (1.4~nm Al) and 0.25 (3~nm Al), values well larger than the one obtained for reference sample C, in the range of the value obtained for the reference sample D with 1.5~nm top
oxidized Al capping instead of 3~nm Pt~\cite{krishnia2025}. This strong FL component manifest the emergence of a peculiar electronic band structure at the Co|Al interface characterized by an OREE and an OML~\cite{Nikolaev_2024_Nanoletters,pezo2024theory}. The behavior of the DL component shown in Fig.~\ref{fig:Efficiency}(b) differs in that $\xi_{DL}$ saturates for Al(1.4) or even decreases for  for Al(3) and its corresponding values (0.07-0.13) are in the range of the reference sample C but not of the one of sample D. One observes that the DL has a maximum at the vicinity of $t_{Co}\simeq~1.2$~nm and we interpret this feature as a pretty fair determination of the finite spin penetration length in Co for SOT based using the spin-hall effects (SHE) of Pt. 

From those data, the ratio between the two SOT components $\zeta=\left(\frac{\xi_{FL}}{\xi_{DL}}\right)$ rises from 0.7 (case of bulk Co) to 2.5 for the series A when $t_{Co}$ is reduced down to 0.5~nm (Fig.~\ref{fig:Efficiency}c). $\zeta (t_{Co})$ follows the exact same trend and dependence for Al(1.4) and Al(3) characterized by a certain enhancement compared to the standard spin-precession mechanism in Co from an SHE source. To compare, we show on Fig.~\ref{fig:Efficiency}(c) the results obtained for Pt|Co|Cu where the light metal Al is replaced by Cu displaying well smaller $\zeta$ ratio (orange points) very close to the situation of total free Rashba component (black dot line). The latter plot has been obtained by only keeping the SHE contribution to the fitting analysis obtained from Pt|Co|Al~3~nm experimental data.

In that SHE picture, predicting SOT requires considering combining precession/relaxation mechanism in Co as well as spin-dependent reflection at interfaces as shown previously~\cite{krishnia2023large}; however without being able to reproduce the experimental trends observed for $\zeta (t_{Co})$. The ensemble of data for series A made of Co|Al interface involving light elements manifests the emergence of an additional contribution: a common OREE inducing FL torque in Co in the series A (Pt capping) and reference D (AlOx capping). Nevertheless, the evolution in the torque properties comparing the between Al 1.4 and Al 3~nm sample, and more generally in the whole $t_{Al}$ series~\cite{krishnia2023large} manifests a gradual variation of the OREE vs. $t_{Al}$ which remains to analyze and explore in details. 

\vspace{0.1in}

\textit{Co|Al with Pt: crossover from OT to SOT}\label{sec:Ptinterf}

\vspace{0.1in}

The OT reported in the previous sections is remarkable because Co|Al is a system made of only light element. A crossover from OT to SOT system may be observable in slightly companion systems whereby a thin Pt heavy metal is gradually inserted between Co and Al. This is the spirit of series B made of interleaved layers made of equivalent few subatomic planes of Pt ($\tilde{\text{Pt}}$), from 0.25 to 3~nm, in order to modify either \textit{(i)} the transverse angular momentum flow (spin-current) or \textit{(i)} the Co|Al electronic band structure through atomic Pt insertion. The growth of the stacks are observed to be smooth and free of large interdiffusion between metallic species preserving thus the interface quality also in the subnanometric $\tilde{\text{Pt}}$ thickness range.

\begin{figure}
\includegraphics[width=0.85\textwidth]{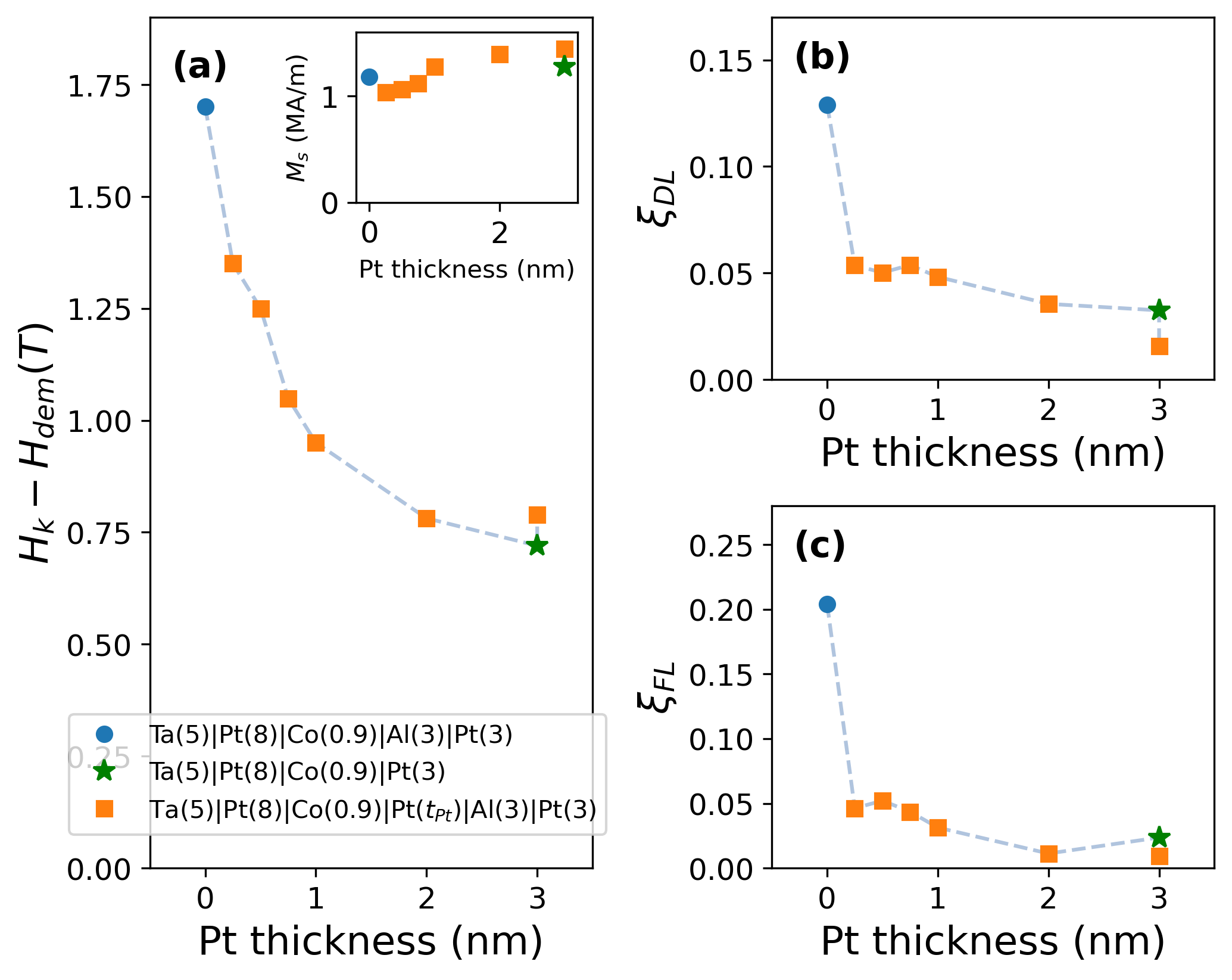}%
\caption{\label{fig:CoPtAl} Anisotropy (a), damping like torque effective field (b) and field like torque effective field (c) measured as a function of the thickness of Pt between Co and Al in $\mathrm{Ta(5)|Pt(8)|Co(0.9)|Pt(}t_{Pt}\mathrm{|Al(3)|Pt(3)}$. In each plot this series of samples is compared to sample without Pt between Co and Al and a sample without Al: $\mathrm{Ta(5)|Pt(8)|Co(0.9)|Pt(3)}$.}
\end{figure}

Fig.~\ref{fig:CoPtAl}(a) displays the resulting effective anisotropy fields $H_K^{eff}=\frac{K_S}{M_S ~t_{Co}}-4\pi M_s>0$ in the (0.25-3)~nm $\tilde{\text{Pt}}$ thickness window ($t_{Co}=0.9~$nm). We observe that the effective anisotropy field $H_k^{eff}$ decreases from a very large value (1.75~T) free of any $\tilde{\text{Pt}}$ intercalation, to 1.35~T for 0.25~nm $\tilde{\text{Pt}}$ down to 0.79~T for 3~nm Pt (Pt(8)|Co(0.9)|Pt(3) system) as reported in our previous work~\cite{krishnia2023large}: Co|Al interface giving $H_k^{eff} =\ 1.72$~T) to the reference case Pt|Co|Pt, $H_k^{eff} =\ 0.72$~T. We emphasize the the very strong PMA anisotropy revealed in Co|Al systems witnesses the occurrence of large interfacial OAM coupled to sizable Rashba and DMI interactions~\cite{barnes2014rashba}. The ensemble of the values of $M_s$ (inset of Fig.~\ref{fig:CoPtAl}(a)) and $K_s$ for the sample series are gathered in the table of the \textcolor{blue}{Suppl. Info. I}.

We now turn to the discussion of DL and FL torque measurements on this series B (Fig.~\ref{fig:CoPtAl}b-c). As a 0.25~nm interleaved $\tilde{\text{Pt}}$ layer is inserted between Co and Al layers, we observe a conjugate decrease of the FL and DL torques by respectively a factor 4 for the FLT ($\xi_{FL}$ drops from 0.22 to 0.055) and factor of 2 for the DLT ($\xi_{DL}$ drops from 0.13 to 0.065), reaching a typical ratio $\zeta=\left(\frac{H_{FL}}{H_{DL}}\right)=0.9$ in the range expected when the Rashba field torque component becomes small (refer to Fig.~2c) in Ref.~\cite{krishnia2023large}). As the interleaved Pt layer gradually increases up to 3~nm, the two torque components gradually decrease up to their expectation value ($\xi_{DL}^{\tilde{\text{Pt}}=3}=0.02$) for asymmetric perfectly layered Pt(8)|Co(0.9)|Pt(3) structures. The experimental ratio for $\tilde{\text{Pt}}=3~$ nm is $\zeta=0.3-0.4$. The ensemble of experimental data and a drop of the FL twice than for the DL (still for 0.25~nm $\tilde{\text{Pt}}$) indicate that the $\tilde{\text{Pt}}$-thickness dependence of the torque may originate from: \textit{i)} enhanced spin-flip scattering in $\tilde{\text{Pt}}$ as explored for the contact layer~\cite{Legrand2016} of for the thin FM layer~\cite{Zhu2023}, \textit{ii)} a rapid disappearance (destruction) of the (orbital) Rashba momentum locking at the Co|$\tilde{\text{Pt}}$|Al interface due to a modification of the band structure \textit{iii)} a rise of the SHE contribution at the top Co|$\tilde{\text{Pt}}$ opposite to the bottom Pt|Co one leading to an almost full torque compensation. 

The first issue \textit{i)} have been previously evoked when discussing AHE experiments. It simply relies on the fact that adding spin-flip at the back-end of Co prevents any electronic reflection having for effect to reduce the effective \textit{s-d} interactions and the DL component by almost a factor two. On the other hand, the reduction of spin-reflection at top Co interface in Co|$\tilde{\text{Pt}}$|Al|Pt systems with $\tilde{\text{Pt}}$ insertion explains a larger DL torque with Al(3) than for Al(1.4) in the series A and a strong reduction of the DL with Pt insertion. Nevertheless, the concomitant increase of the FL torque from Al(1.4) to Al(3) in the series A mainly originate from the increase of the orbital Rashba interaction possibly associated to the lattice strain as shown by DFT calculations~\cite{krishnia2025}.

\section{Discussions}

\subsection{Semi-classical SOT modeling from pure SHE}

In this section, we first analyze our data from a pure SHE model and show that such assumption of controlling the spin-flip rate~\cite{Legrand2016,Zhu2023}, here by introducing $\tilde{Pt}$, cannot account for the results. To this end, We consider the Pt|Co|$\tilde{Pt}$|Al system and consider the injection of a SHE current from Pt into Co|$\tilde{Pt}|$Al, the latter being characterized by a spin-flip conductance or spin-flip rate $\mathcal{G}_{sf}$ which can be tuned by the $\tilde{Pt}$ content. $\mathcal{G}_{sf}$ is then expected to vary from 0 (no $\tilde{Pt}$) to very large vales when increasing the $\tilde{Pt}$ content.

In order to quantify the DL and FL torque efficiencies from SHE, we have to consider the in-depth absorption of the transverse SHE spin-current into Co taking into account precession and spin-flip processes, and comparing the A and B sample series. By referring to the concepts described in the (\textcolor{blue}{Suppl. Info. part III}), one obtains the effective SHE SOT current $\boldsymbol{\hat{\Gamma}}_{DL,FL}^{SHE}=\xi_{DL,FL}\left(\sigma_{SHC}\mathbf{E}\right)$ with $\xi_{DL,FL}$ the SOT efficiency:

%\begin{widetext}
%\begin{equation}
%\begin{aligned}
%\frac{\xi_{DL,FL}^{SHE}}{\left(\theta_{SHE}^*\right)}\simeq \frac{r_s^{Pt} \left[1+\mathcal{G}_{sf}r_F^*\tanh{\left(\frac{t_F^*}{2}\right)}\right]}{\left[r_s^{Pt}+\mathcal{G}_{sf}\left(r_F^*\right)^2\right]+r_F^*\left(1+\mathcal{G}_{sf}r_s^{Pt}\right)\coth{\left(t_F^*\right)}}
%\end{aligned}
%\label{she_current}
%\end{equation}
%\end{widetext}

\begin{widetext}
\begin{equation}
\begin{aligned}
\frac{\xi_{DL,FL}^{SHE}}{\left(\theta_{SHE}^*\right)}\simeq \frac{\left(\mathcal{G}_{\uparrow\downarrow}^{Pt|Co}r_s^{Pt}\right) \left[\mathcal{G}_{\uparrow\downarrow}^{Pt|Co}+\mathcal{G}_{sf}\tanh{\left(\frac{t_F^*}{2}\right)}\right]}{\mathcal{G}_{sf}+\mathcal{G}_{\uparrow\downarrow}^{Pt|Co}\left[\mathcal{G}_{\uparrow\downarrow}^{Pt|Co}r_s^{Pt}+\left(1+\mathcal{G}_{sf}r_s^{Pt}\right)\coth{\left(t_F^*\right)}\right]}
\end{aligned}
\label{she_current}
\end{equation}
\end{widetext}
where the DL (FL) torque is represented by the respective Real and Imaginary parts of $\chi_{DL,FL}$. In this equation~[\ref{she_current}], $\theta_{SHE}^*=\theta_{SHE}\exp{\left(-\delta\right)}$ is the effective spin-Hall angle accounting for the spin-memory loss. From AHE $\delta\simeq 0.4$ at the Pt|Co interface.  We have introduced the spin-mixing conductance of the bottom Pt8|Co interface $\mathcal{G}_{\uparrow \downarrow}$ as a generalization of the effective spin-resistance $\mathcal{G}_{\uparrow \downarrow}^{Pt|Co}\simeq \left(r_F^*\right)^{-1}$ for the transverse spin angular momentum. It is defined for a thick FM layer whereas the $\tanh\left(t_F^*\right)$ and $\coth\left(t_F^*\right)$ dependence involve such thickness dependence where $t_F^*=\left(\frac{t_F}{\lambda_F^*}\right)$ is the reduced Co thickness ($t_F$) in unit of the complex effective diffusion length ($\lambda_F^*$) owing to the precession process.
$r_s^{Pt}$ stands for the spin-flip resistance of the 8~nm bottom Pt. 

We consider several limits:

\begin{itemize}
    \item $t_F^*\rightarrow 0: \\\left(\frac{\xi}{\theta_{SHE}^*}\right)\rightarrow \left(\frac{\mathcal{G}_{\uparrow\downarrow}^{Pt|Co}r_s^{Pt}t_F^*}{1+\mathcal{G}_{sf}r_s^{Pt}}\right)=\left(\frac{\mathcal{G}_{\uparrow\downarrow}^{Pt|Co}t_F}{\mathcal{G}_{sf}\lambda_{F}^*}\right)\left(\frac{\mathcal{G}_{sf}r_s^{Pt}}{1+\mathcal{G}_{sf}r_s^{Pt}}\right)$ showing that the interaction between spin-current and local magnetization varies linearly with $t_F$. This formula indicates that the integrated SOT current in the thin Co ferromagnet is proportional to the product of the overall spin-injection efficiency (free of bulk Co at the limit of zero thickness) $\left(\frac{\mathcal{G}_{sf}r_s^{Pt}}{1+\mathcal{G}_{sf}r_s^{Pt}}\right)$ $\otimes$ the Co thickness \textit{via} the ratio of the probability of spin-flip $\left(\frac{\mathcal{G}_{\uparrow\downarrow}^{Pt|Co}t_F}{\mathcal{G}_{sf}\lambda_{F}^*}\right)$ in Co ($\propto \mathcal{G}_{\uparrow\downarrow}^{Pt|Co}\frac{t_F}{\lambda_F^*}$) over the spin-flip rate in the outer $\tilde{Pt}$ ($\propto \mathcal{G}_{sf}$).
    \item $t_F^*\rightarrow\infty:\\\left(\frac{\xi}{\theta_{SHE}^*}\right)\rightarrow \left(\frac{\mathcal{G}_{\uparrow\downarrow}^{Pt|Co}r_s^{Pt}}{1+\mathcal{G}_{\uparrow\downarrow}^{Pt|Co}r_s^{Pt}}\right)$ showing that the interaction is independent on the outer interface parametrized by $\mathcal{G}_{sf}$. The complex character of $r_F^*$ implies a complex effective electronic transmission/reflection scaled by the spin-mixing conductance.
    \item 
    $\mathcal{G}_{sf}\rightarrow 0:\\\left(\frac{\xi}{\theta_{SHE}^*}\right)\rightarrow \left(\frac{\mathcal{G}_{\uparrow\downarrow}^{Pt|Co}r_s^{Pt}}{\mathcal{G}_{\uparrow\downarrow}^{Pt|Co}r_s^{Pt}+\coth{\left(t_F^*\right)}}\right)$. This corresponds to a perfect spin reflection at the outer boundary (case of Co|Al outer boundary). 
    \item 
    $\mathcal{G}_{sf}\rightarrow \infty:\\\left(\frac{\xi}{\theta_{SHE}^*}\right)\rightarrow \frac{\left(\mathcal{G}_{\uparrow\downarrow}^{Pt|Co}r_s^{Pt}\right)\tanh{\left(\frac{t_F^*}{2}\right)}}{1+\left(\mathcal{G}_{\uparrow\downarrow}^{Pt|Co}r_s^{Pt}\right)\coth{\left(t_F^*\right)}}$ corresponding to a perfect spin sink at the outer boundary (case of Co|$\tilde{Pt}$ outer boundary).
\end{itemize}
In the last two limits, one recovers the previous formula in the limit of thick 'Co' because then independent of the outer boundary conditions.

\vspace{0.1in}

We now describe the general situation of varying $\mathcal{G}_{sf}$ from small values for low $\tilde{Pt}$ content to large values corresponding to high $\tilde{Pt}$  content for Pt|Co|$\tilde{Pt}$|Al systems. The result of our fitting procedures for the OT/SOT gives a characteristic Larmor length of $\lambda_J=\left(\frac{\hbar v_F}{\Delta_{xc}}\right)=0.65\pm 0.05$~nm, a transverse decoherence length $\lambda_\Delta=1.2\pm 0.1$~nm, $r_s^{Pt}=\sqrt{\frac{3\tau_{sf}}{\tau_p}}\simeq 2$ (in unit of $G_{Sh}^{-1}$, (\textcolor{blue}{Suppl. Info. part III}), $\left(G_{\uparrow\downarrow}^{Pt|Co}\right)^{-1}\simeq r_F^*=\rho_F\times \lambda_F^*\approx 3-0.07i ~\left(G_{sh}\right)^{-1}$ and typical ferromagnetic thickness $t_F<1.5$~nm.

\vspace{0.1in}

It turns out that on varying $\mathcal{G}_{sf}$ and considering our physical parameters, $\xi_{DL}$ decreases typically from 0.06 to 0.035 for $t_F=0.8$~nm as the rate of spin-flip due to $\tilde{Pt}$ increases (from Co|Al to Co|$\tilde{Pt}$|Al). This correspond to a drop of about 40\% as observed in our measurements. Corollary, from our model $\xi_{FL}$ is observed to increase from 0.01 to 0.035 on increasing $G_{sf}$. It results that intercalating $\tilde{Pt}$ content in-between Co|Al would have for effect to enhance much the ratio from $\zeta=\left(\frac{\xi_{FL}}{\xi_{DL}}\right)\simeq 0.2$ to $\zeta=1$. If the behavior of $\xi_{DL}$ seems to be in agreement with our data, that is a decrease by about a factor of two, the experimental behavior of $\xi_{FL}$ associated to a drop by a factor 6 for 0.25~nm $\tilde{Pt}$ does not fit with this SHE model. The same qualitative conclusion, a decrease of $\chi_{DL}$ and an increase of $\chi_{FL}$ with $\mathcal{G}_{sf}$ applies for the FM thickness in the range 0.5-1.5~nm.

\subsection{Insights from DFT on the OREE mechanism}

The strong FL SOT and OT evidenced at Co|Al interfaces cannot then be explained neither by standard SOT mechanisms nor by processes already explored involving oxidized light metals  and largely investigated in a recent literature as using \textit{e.g.} CuO~\cite{Kim_2021_nontrivial,ding2020harnessing,Sala_2022_3d5d4f,kim2023oxide,krishnia2024quantifying,bony2025quantitative,xu2025alternative}, metallic Cu~\cite{yi2025}, Zr~\cite{Ando2023} or CoO~\cite{wang2024,Ding2025}. In those orbital FM|LM systems, an heavy layer like Pt~\cite{ding2020harnessing,krishnia2024quantifying,bony2025quantitative,xu2025alternative} or rare-earth materials (Gd, Tb)~\cite{Sala_2022_3d5d4f,gambardella2024} is generally used as interleaved spacer (FM|HM|LM) to introduce a second additive spin channel source and to enhance spin-injection efficiency and subsequent magnetic torques.

\begin{figure}[h!]
\centering
%\hspace*{-0.55cm}
\includegraphics[scale=0.45]{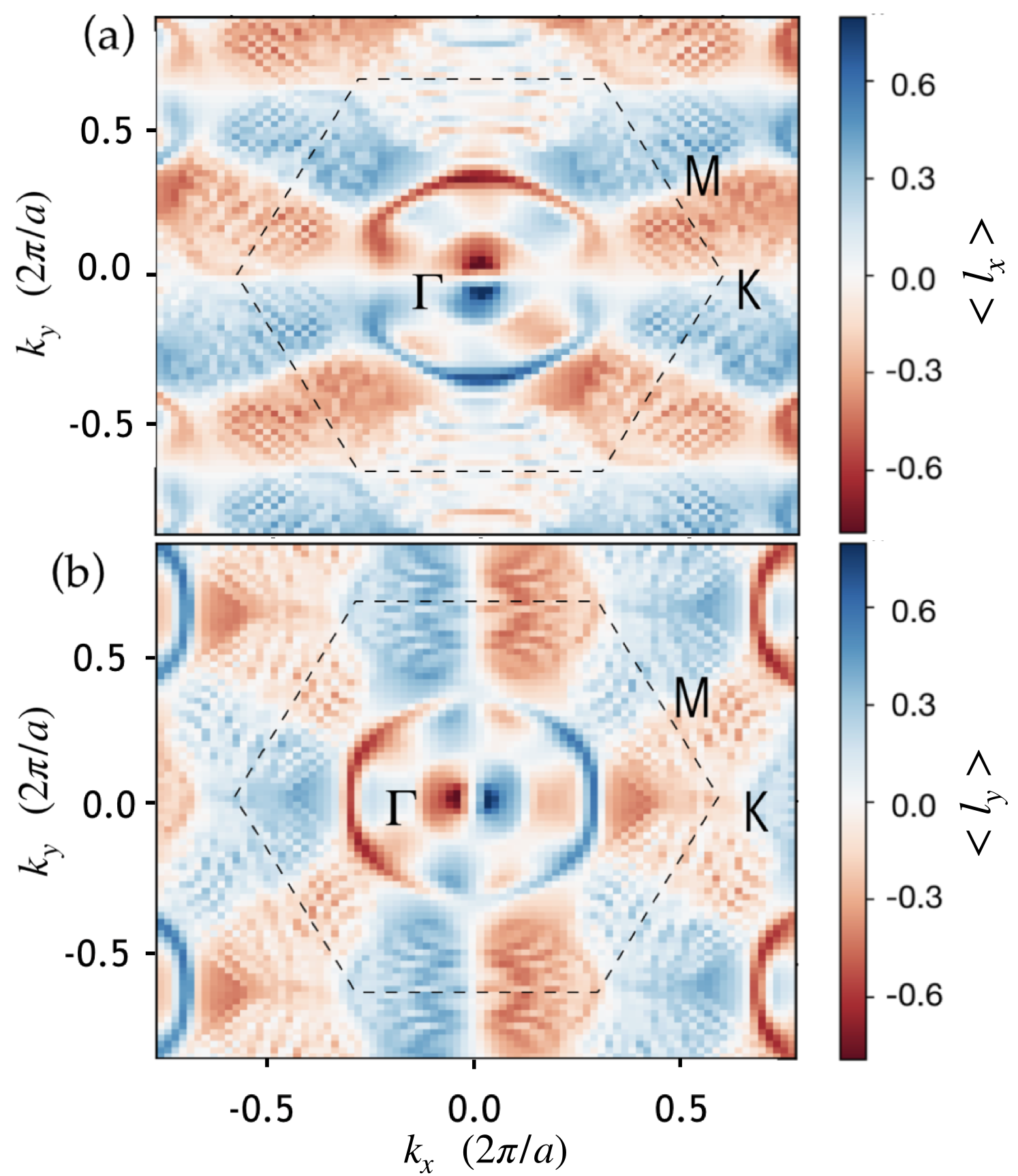}
\caption{Expectation values of the $\braket{\hat{L}_x}$ (left panel) and $\braket{\hat{L}_y}$ (right panel) projected OAM onto the Fermi surface of Co(12)|Al(12) interfaces displaying an orbital moment transverse to the direction of the electronic wavevector $k$: the OML.}
\label{Fig:lx_texture_coal}
\end{figure}

In order to gain insights onto the actual underlying physics, first principle calculations, performed by DFT appears as a powerful tool to distinguish spin and orbital contributions. The following calculations allow to draw SOT and OT following three consecutive steps: \textit{(i)} the use of DFT yields the spin and orbital texture (SML and OML), \textit{(ii)} the implementation of the Kubo formalism for quantifying the linear OREE response under the external electric field, \textit{(iii)} the calculation of the two torque components (or torkance) acting onto the magnetization.

\vspace{0.1in}

\textit{Emergence of the OML}

\vspace{0.1in}

DFT combined with Wannierization techniques was previously used to benchmark Co(12)|Al(12) to Co(12)|Cu(12) systems (concerning DFT, the number in parenthesis stems for the thickness in atomic layer units) as we learned from previous studies that Co|Cu systems do not present any torque amplification unlike the case of Co|Al~\cite{Nikolaev_2024_Nanoletters}. Regarding the interface between Co and light metals as Cu or Al, while a weak coupling is found in the case of Cu, it is particularly strong for Co|Al, especially at the $K$ point of the Brillouin zone edge.
Notably, at the $K-$point, we could observe electronic states in Co|Al that do not appear in Co|Cu. The most striking outcome of the hybridization onto Co(12)|Al(12) is the emergence of the OML texture in the 2D Brillouin zone. We clearly observe at the level of the Co$_1$ layer (first Co plane in contact with Al) a strong OML, absent for Co(12)|Cu(12).

\begin{figure}[h!]
    \centering
\includegraphics[width=0.85\textwidth]{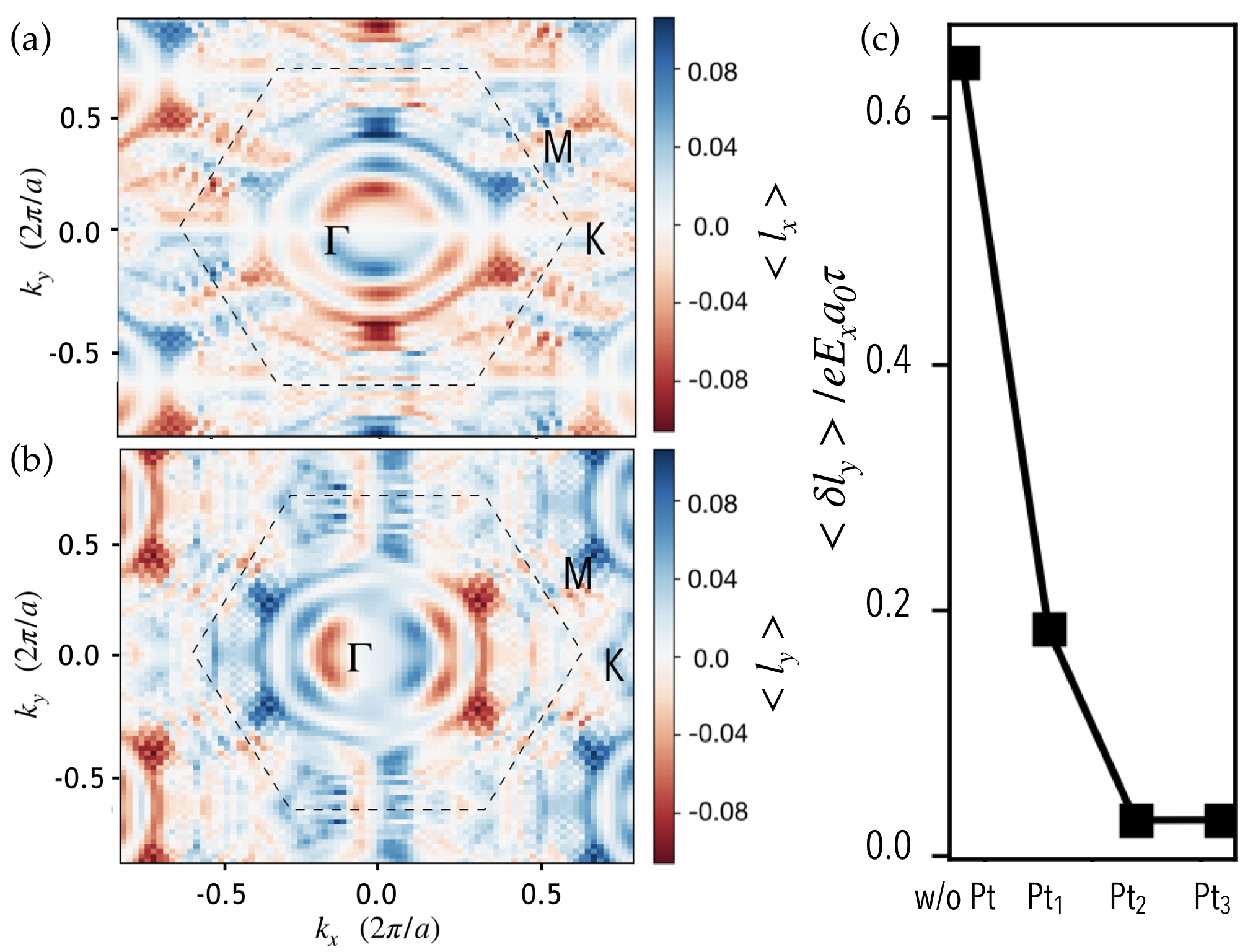}
    \caption{Expected values of (a) $\braket{l_x}$ and (b) $\braket{l_y}$ projected over the Brillouin zone in Al|Pt(1)|Co. c) Integrated orbital moment density (accumulation) in response over the full Co thickness in Co(12)|Al(12) and CO(12)|Pt(1,2,3)|Al interfaces. \textit{numbers refers to the stacked atomic plane}.}
    \label{fig:coptal_orbit}
\end{figure}

From our DFT treatment presently employed (\textcolor{blue}{methods}), we recover the ensemble of the aforementioned features related to the occurrence of OML in Co|Al systems unlike the case of Co|Cu~\cite{pezo2025anatomy}.
This is particularly exemplified in Fig.~\ref{fig:coptal_orbit}(a-b) clearly displaying the locking of the orbital moment expectation value, $\braket{\hat{L}_x}$ $\braket{\hat{L}_y}$ (the horizontal vs. vertical directions stands for the respective $k_x$ and $k_y$ directions). One clearly observes that the $\braket{\hat{L}_x}$ ($\braket{\hat{L}_y}$) orbital moment component is vanishing along $\hat{x}$ ($\hat{y}$). Note that the amplitude of $\braket{\hat{L}}$ onto the Fermi surface reaches 0.8~$\mu_B$, a value which appears even larger than the values observed onto topological surface states (TSS)~\cite{pezo2024theory}. Moreover, in agreement with Ref.~\cite{Nikolaev_2024_Nanoletters}, we demonstrate the absence of significant SML feature in those systems (\textcolor{blue}{Suppl. Info. part III}) anticipating thus the possibility of a pure OREE response responsible for the magnetic torques.

We turn to Co(12)|Pt(1,2,3)|Al(12) systems. Fig.~\ref{fig:coptal_orbit}(a-b) display the expected values of $\braket{\hat{L}_x}$ and $\braket{\hat{L}_y}$, the orbital texture when a single atomic plane of Pt is inserted between Co(12) and Al(11). One observes a remarkable drop of the $\braket{\hat{L}_{x,y}}$ amplitude corresponding to a severe loss of the OML down to range of values less than 0.1~$\mu_B$. In full agreement with our experimental data, this demonstrates that the insertion of Pt, even at the lower atomic limit, results in the disappearance of the OML built from Co|Al. This also reinforces the major role of the \textit{3p-3d} hybridization and orbital bonding at the Co|Al bare interface promoting the OML. 

\vspace{0.1in}

\textit{OREE linear response}

\vspace{0.1in}

We now focus on the linear response theory of the orbital Rashba-Edelstein (OREE) response $\chi_{xy}^L$ with $\braket{L}=\chi_{xy}^{\hat{L}}~\mathbf{E}$ obtained from DFT using the Kubo response theory~\cite{go2020theory}. The corresponding components are given by the respective \textit{intraband} vs. \textit{interband} terms formulae~\cite{pezo2025anatomy}:

\begin{widetext}
\begin{flalign}
    \chi_{\alpha \beta}^{L,intra} &= -\left(\frac{ i e \tau}{N_k}\right) \sum_{\mathbf{k}}^{n=\text{occ}} \left(\frac{\partial f_{n\mathbf{k}}}{\partial \epsilon_{n\mathbf{k}}}\right) \bra{u_{n\mathbf{k}}} \hat{O}_\alpha \ket{u_{n\mathbf{k}}} \bra{ u_{n\mathbf{k}}} \hat{v}_{\beta \mathbf{k}} \ket{ u_{n\mathbf{k}}} \label{eq:intraL} \\
    \chi_{\alpha \beta}^{L,inter} &= \left(\frac{e \hbar}{2\pi N_k}\right) \sum_{\mathbf{k}, m\neq n}^{n=\text{occ}} \left(f_{n\mathbf{k}} - f_{m\mathbf{k}} \right)\frac{ \bra{u_{n\mathbf{k}}} \hat{O}_\alpha \ket{u_{m\mathbf{k}}} \bra{u_{m\mathbf{k}}} \hat{v}_{\beta \mathbf{k}} \ket{u_{n\mathbf{k}}}}{\left(\epsilon_{n\mathbf{k}} - \epsilon_{m\mathbf{k}}\right) \left( \epsilon_{n\mathbf{k}} - \epsilon_{m\mathbf{k}} - i \hbar \tau^{-1} \right)} 
    \label{eq:linear_response}
\end{flalign}
\end{widetext}
with $\hat{O}_\alpha=\hat{L}_\alpha$ is a generator quantum mechanical operator, presently the $\alpha$-component of the orbital moment $\hat{L}$ operator. $\hat{v}_{\beta\mathbf{k}}$ is the electron velocity along the direction $\beta$ at wavevector $\mathbf{k}$. We have introduced a finite value of the Fermi broadening energy $\Gamma_F=0.075$~meV and $\varepsilon_F$ is the energy Fermi level. This value of $\Gamma_F=0.075$ corresponds to a typical scattering relaxation time $\tau=\frac{\hbar}{2\gamma_F}\simeq$10~fs matching with the scattering time in the bulk Co and corresponding average mean-free-path $\lambda_{Co}$ of about 5~nm. $N_k$ is the number of sampling point in the reciprocal space. Further on, we will use equivalent definition when dealing with the torque $\mathbf{T}_\alpha$ for the torkance calculation. 

From simple symmetry arguments, the \textit{intraband} (respectively \textit{interband}) contribution is expected to cancel for the $\chi_{xy}^{L_x}$ (respectively $\chi_{xy}^{L_y}$) DL component if the Time reversal symmetry (TRS) is invoked (case of non-magnetic structures)~\cite{Johansson2024theory}. The response of Co|Al to an electric field in terms of both spin and orbital angular accumulations are computed in Refs.~\cite{Nikolaev_2024_Nanoletters,pezo2025anatomy}. Despite slight quantitative differences, the two studies draw the same qualitative conclusions: a large orbital response occurs in the very first atomic plane of Co in direct contact with Al. Both calculations find a much larger susceptibility for the orbital accumulation than for the spin counterpart.

\begin{table}[h!]
    \centering
    \begin{tabular}{c|c|c}
            &    Ref.~\cite{pezo2025anatomy}
            (DFT)&   Ref.~\cite{Nikolaev_2024_Nanoletters}(Wannier) \\
        \hline
        lattice param. (\AA) &   2.94    &   2.86   \\
        \hline
        $\chi^{L,intra}_{xy}$ &   5.94    &   $\simeq 2.0$ \\
        $\chi^{L,inter}_{xy}$ &   0.20    &   \\
        $\chi^{S,intra}_{xy}$ &   0.12    &   $\simeq 0.15$ \\
        $\chi^{S,inter}_{xy}$ &   -0.002    &   \\
    \end{tabular}
    \caption{\textbf{Co12|Al12}: Comparison between $\chi_{xy}^{L,S}$ orbital- and spin- REE obtained from DFT and Wannier calculations performed in Refs.~\cite{pezo2025anatomy, Nikolaev_2024_Nanoletters} corresponding to two different lattice parameters. OREE response $\chi$ are given in unit of $10^{-10}~\hbar/$~(V/m) per atom. The orbital polarization of (orbital accumulation density) induced by OREE is then $\hat{\mu}_L=\chi_{xy}^{L} \mathbf{E}~\left(10^{-10}~\hbar\right)/$(V/m.atom). }
    \label{tab:dftvalues}
\end{table}

\vspace{0.1in}

Still, one difference is worth noting. In Nikolaev et al.~\cite{Nikolaev_2024_Nanoletters}, the response in spin accumulation is approximately ten times smaller than the orbital response, whereas in Pezo et al.~\cite{pezo2025anatomy}, we find that it is smaller by a factor close to 50. Regarding the SOT generation, we should emphasize that the torques exerted by the orbital accumulation within Co are mediated by SOC which is typically of the order of 0.07~eV in Co. Consequently, Ref.~\cite{Nikolaev_2024_Nanoletters} does not exclude the role of Rashba induced spin accumulation in SOT while Ref.~\cite{pezo2025anatomy} fully rules out such possibility. %This difference between the two works at this point might stem from an additional approximation introduced by the Wannierization technique being used in Ref.~\cite{nikolaev2024large} (which is presently not the case in Ref.~\cite{pezo2025anatomy}). This approximation reduces the number of bands used for computing sums \ref{eq:intraL} and \ref{eq:interL} by focusing on electronic states and bands close to the Fermi level.

\vspace{0.1in}

Table~\ref{table_xi} gathers the OREE response $\chi_{xy}^L$ calculated onto the first Co atomic layer in contact with either Al in Co|Al. Note that for Co|Al, the value of $\chi_{xy}^L$ is in very good agreement with the value expected from the experimental torque if one compares with Eq.~\ref{chi_ordermagnitude} comforting thus our assumption of OREE. The calculated torkance are presented in the following section

\vspace{0.1in}

\textit{Torkance linear response}

\vspace{0.1in}

Hereafter, we conducted simulations of the OT from first-principle techniques using the linear response theory obtained from the Kubo's theory~\cite{freimuth2014}. The full Hamiltonian can be decomposed into the kinetic part, $\hat{H}_K$, a SOC term, $\hat{H}_{SOC}=\xi_{SOC}\left(\hat{\nabla} V(r)\times \hat{p}\right)\cdot \hat{\mathbf{\sigma}}$, small for a 3\textit{d} magnet, but however mandatory for the torque process, plus an exchange-correlation part $\hat{H}_{xc}=-\mu_B\mathbf{B}_{xc}\cdot\mathbf{\hat{\sigma}}$, where the exchange field $\mathbf{B}_{xc}$ accounts for the difference between the effective Kohn-Sham potentials for minority and majority quasi-particles in Co. By applying an electric field $\mathbf{E}$, an out of equilibrium change in the magnetization $\delta \mathbf{M}$ enables the modification of the exchange following $\delta \mathbf{B}_{xc}=B_{xc}\delta \mathbf{M}/M$. Such a dynamical effect leads to the action of the torque $\hat{\mathbf{T}}$ onto $\mathbf{M}$ within the unit cell given by $\boldsymbol{\hat{\Gamma}}_{SOT}=\int dV~ \mathbf{M} \times \delta \mathbf{B}_{xc}=-\int dV~ \mathbf{B}_{xc} \times \delta \mathbf{M}$.

\vspace{0.1in}

Within the linear response, the torque $\boldsymbol{\hat{\Gamma}}_{SOT}$ can be written in terms of the torkance $\mathbf{t}$ such that $\hat{\Gamma}_{SOT}=\mathbf{t}\mathbf{E}$~\cite{belashchenko1,belashchenko2,go2020theory}. Such tensor $\mathbf{t}$ can be decomposed into a respective \textit{interband} and \textit{intraband} terms $\mathbf{t}_{ij}=\mathbf{t}^{\text{Inter}}_{ij}+\mathbf{t}^{\text{Intra}}_{ij}$ according to the previous equation~\ref{eq:linear_response} after having replaced the general operator $\hat{O}_\alpha$ by the torque operator $\hat{\mathbf{T}}_\alpha=\frac{-i}{\hbar}[\hat{H},\hat{\mathbf{s}}]_\alpha$ = $-\left[\mu_B \hat{\mathbf{\sigma}} \times \mathbf{B}_{xc}\right]_\alpha$. The exchange contribution to the Hamiltonian represents the main term owing in the case of a small SOC and is proportional to $\hat{\Omega}^{xc}=\Omega^{xc} \hat{\mathbf{m}}$ and $\Omega^{xc}=\frac{1}{2\mu_B}[V^{eff}_{min}(\vec{r})-V^{eff}_{maj}(\vec{r})]$ is the exchange field.  
We took this part as the values of individual atomic magnetic moments after the self-consistent calculation.

%\begin{flalign}
%    \mathbf{t}_{\beta\alpha}^{\text{Inter}}&=\left(\frac{e\hbar}{\pi N_k}\right)\sum_{\mathbf{k}, m\neq n}^{n=\text{occ}}\frac{\operatorname{Im}\big [\braket{\psi_{\mathbf{k}n}|\mathbf{\hat{T}}_\alpha|\psi_{\mathbf{k}m}}\braket{\psi_{\mathbf{k}m}|\hat{v}_\beta|\psi_{\mathbf{k}n}}\big]}{(\varepsilon_m-\varepsilon_n)^2}\\
%    \mathbf{t}_{\beta \alpha}^{\text{Intra}}&=\left(\frac{e\hbar}{2 \Gamma N_k}\right)\sum_{\mathbf{k}}^{n=\text{occ}} \operatorname{Re}\big [\braket{\psi_{\mathbf{k}n}|\mathbf{\hat{T}}_\alpha|\psi_{\mathbf{k}n}}\braket{\psi_{\mathbf{k}n}|\hat{v}_\beta|\psi_{\mathbf{k}n}}\big] \delta(\varepsilon_F-\varepsilon_n)
%\end{flalign}

\vspace{0.1in}

From symmetry arguments, the partition into \textit{interband} and \textit{intraband} has to be assigned to mainly damping-like torque or DLT ($\mathbf{T}_{xy}$) and field-like torque or FLT ($\mathbf{T}_{xx}$) components respectively~\cite{go2020theory,Hayashi2023} despite the lack of a mirror symmetry due to the hexagonal stacking may induce small mixing terms between FL and DL terms~\cite{Nikolaev_2024_Nanoletters} (see also Table~\ref{table_xi} comparing the orbital polarization $\braket{\hat{L}_y}$ for respective \textit{intraband} vs. \textit{interband} terms). The torkance tensor was computed in our Ref.~\cite{pezo2025anatomy} assuming that $t_{xx}^{inter}$ and $t_{xy}^{intra}$ are zero. The integrated torque acting on a volume unit of Co (per unit surface) are given in unit of ($ea_0$) ($a_0$ is the Bohr radius). The corresponding effective SOT magnetic fields $\mathbf{B}_{SOT}$ are then given by: $\mathbf{B}_{SOT}=\left(\frac{e~a_0}{M_{s,at}^{Co}~N_{Co}}\right)~t_{xx}~\mathbf{E}$ with $M_{s,at}$ the atomic magnetization and $N_{Co}$ the number of Co planes in the thickness. A value of $t_{xx}=1$ gives $\mathbf{B}_{SOT}=4.5$~mT for $\mathbf{E}=2.5\times 10^4$~V/m (current density in Pt of $J_c=10^{11}$~A/m$^2$). The average over the whole Co thickness presents a factor of 10 between the component associated to field like $t_{xx} = 0.37\ e a_0$ and the damping like component $t_{xy} = -0.04\ e a_0$. This confirms that SOT emerging at Co|Al interface are prominently of field-like symmetry.

\begin{widetext}
\begin{center}
\begin{table}[h!]
\begin{ruledtabular}
 \begin{tabular}{cccc} 
  Co|Al &$\chi_l^{intra}$$ (10^{-10}\hbar$ $m/(V.atom))$ & $\chi_l^{inter}$$ (10^{-10}\hbar$ $m/(V.atom))$& $t_{xx}$$ (ea_0)$ \\ 
 \hline
 w/o Pt  & 5.94 & 0.20 &  0.37 \\ 
 1 Pt    & 1.56 & 0.18 & -0.08 \\
 2 Pt    & 0.52 & 0.22 &  0.10 \\
 3 Pt    & 0.44 & 0.16 &  0.15 \\ 
 \end{tabular}
 \end{ruledtabular}
 \caption{\label{table:transport_coal} \textbf{Co(12)|Al(12) and Co(12)|Pt(1,2,3)|Al(12)} (numbers in unit of uc): $\braket{\hat{l}_y}$ OREE response at the Co1 plane interfacing Al or Pt. Values are given per atom. Integrated \textit{intraband} $t_{xx}$ (FL) and \textit{interband} $t_{xy}$ (DL) torkance components for Co/Al interfaces without/with Pt insertion. OREE response $\chi$ are given in unit of $10^{-10}\hbar$~m/V (per atom). The orbital polarization of (orbital accumulation density) induced by OREE is then $\hat{\mu}_L=\chi_{xy}^{L} \mathbf{E}$.}
 \label{table_xi}
\end{table}
\end{center}
\end{widetext}

\subsection{OT driven magnetization switching}

\begin{figure}[h!]
\includegraphics[width=0.85\textwidth]{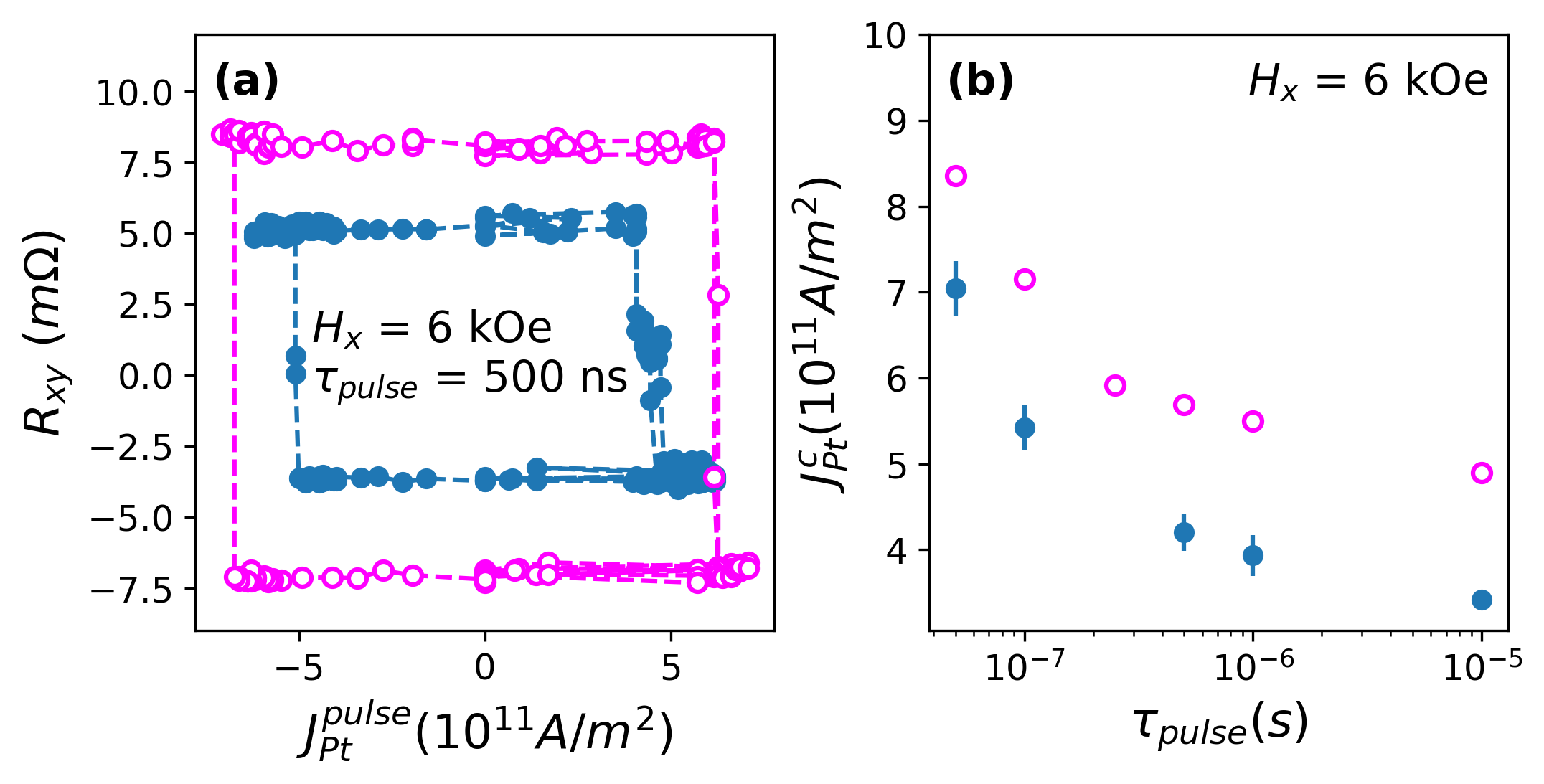}%
\caption{\label{fig:Switch} (a) Magnetization switching cycles vs. the current density $J_c^{Pt}$ injected into the bottom Pt8 observed at room temperature for 500~ns long pulses in Ta5|Pt8|Co0.9|Al1.4|Pt3 (open pink dots) and Ta5|Pt8|Co0.9|Al3|Pt3 (filled blue dots). An in-plane field $H_x$ = 0.6~T is applied. 
(b) Critical current density in Pt for magnetization switching Vs. the current pulse width. The error bars are based on the standard deviation of 20 switching cycles with alternating sign of the in-plane field. An $H_x$ = 0.6~T in-plane field is applied along the current line in the Hall cross bar.}
\end{figure}

In this last section, we now discuss of orbital torque assisted (OT-assisted) magnetization switching performed on $250\times350$~nm$^2$ nanopillars patterned on submicronic Hall cross bars. The measurement protocol are given in the  (\textcolor{blue}{methods}).  We then directly compare the Pt8|Co0.9|Al1.4|Pt3 and Pt8|Co0.9|Al3|Pt3 -based structures from the series A and different in their interleaved Al thickness. It is important to notice that the Pt8|Co0.9|Al3|Pt3 sample (3~nm Al) possesses a stronger perpendicular magnetic anisotropy (Fig.~\ref{fig:Switch}a), 1.72~T vs. 1.5~T for the 1.4~nm Al). Also the very large PMA anisotropy requires a large in-plane external magnetic field $H_x$= 0.6~T to observe a full observable magnetization switching within our experimental conditions. Nevertheless, despite its larger PMA, the critical switching currents were measured significantly smaller on the 3~nm Al sample compared to the 1.4~nm Al~\ref{fig:Switch}(a) thus indicating a significantly larger effective DL torque.
Moreover, we display on \ref{fig:Switch}(b) the resulting values of the critical current density for switching $J_{cr}^{Pt}$ acquired for different current pulse width from 10~$\mu$s down to 50~ns. The reduction of the critical current for the 3~nm Al sample is always observed, from 35\% (100~$\mu$s) to , 15\% (50~ns), without any renormalization made on the difference in their PMA.

About the switching process in the structures investigated, SOT switching has been shown to occur through domain nucleation at an edge of the magnetic pillar followed by the propagation of the subsequent domain wall across the whole magnetic element~\cite{baumgartner2017spatially}. The nucleation is caused by the constructive addition of DMI, in plane field, DL and FL effective fields. However, under DMI, FL tends to be detrimental to the domain wall propagation throughout the magnetic disc. Therefore, the switching critical current reduction in the 3~nm Al sample compared to 1.4~nm Al may not be assigned itself to the larger FL torque promoted by 3~nm Al.

However, from our modeling, OREE does not induce only FL SOT, even though predominant in our Al-based systems, but also a significant part of the DL enhancement observed in our systems. Such additional DL component originates also from the OML and subsequent orbital polarization characterizing the electronic states at the Co|Al interface, and diffusing out by part into the Co layer from the top interface. Such diffusion process is still allowed from the finite electronic escape time out-of the Rashba states towards the metallic reservoir~\cite{sanchez2013spin}.

In order to quantify the DL enhancement vis OREE, we have adapted our model to our experimental results Fig.~\ref{fig:Efficiency} by considering a certain probability of \textit{escape process} parametrized by an interfacial electronic transmission $T_{i1}$. Keeping the other parameter fixed, we are then able to explain the evolution of the two torques components for the two samples. This enables us to quantify the proportions of each SOT component induced by OREE and SHE. We list here below (see \textcolor{blue}{Suppl. Info. part III}):

\begin{itemize}
\item The SHE arising from the bottom Pt|Co layer is parameterized by a bare spin-angle angle of $\theta_{SHE}=0.22\pm 0.02$ and a spin-memory loss coefficient $\delta=0.4$. Such SHE spin current is also subject to spin-backflow at this interface. The efficiency of the torques calculated for $t_{Co}=0.9~$nm are respectively $\xi_{DL}^{SHE}=0.045$ and $\xi_{FL}^{SHE}=0.03$.

\vspace{0.1in}

\item The Ta(5)|Pt(8)|Co(0.9)|Al(1.4)|Pt(3) sample experiments OREE giving rise to \textit{additional torque} components $\xi_{DL}^{OREE}=0.025$ and $\xi_{FL}^{OREE}=0.075$. The FLT from OREE represents 75\% of this additional OREE torque (DLT represents 25\%) showing the major role of REE to generate the FLT. This ratio is parameterized by the $T_{iL}\simeq$0.2 transmission coefficient (coupling) from the Rashba virtual layer to bulk Co states as discussed by Rojas-Sanchez et al.~\cite{sanchez2013spin}. The OREE represents then 44\% of the total DLT and 75\% of the total FLT.

\vspace{0.1in}

\item The OREE on Ta(5)|Pt(8)|Co(0.9)|Al(3)|Pt(3) gives rise to additional torque components $\zeta_{DL}^{OREE}=0.07$ and $\xi_{FL}^{OREE}=0.21$. The Field-like torque by OREE then represents 75\% of the total torque (DLT by OREE is only 25\%). This ratio is parameterized by a coupling $T_{iL}\simeq$~0.1 (transmission coefficient) from the Rashba to bulk Co states (see Ref.~\cite{sanchez2013spin}).

\vspace{0.1in}

\item Our calculations reproduce then well the magnitude of the experimental torques reported here. The final conclusion is that, Ta(5)|Pt(8)|Co(0.9)|Al(1.4)|Pt(3) and Ta(5)|Pt(8)|Co(0.9)|Al(3)|Pt(3) samples, the total DLT efficiency increases from $\xi_{DL}=0.075$ to $\xi_{DL}=0.12$ whereas for the FLT , the increase goes from $\xi_{FL}=0.11$ to $\xi_{FL}=0.23$.
\end{itemize}

The reduced critical switching current observed in the Co|Al sample can thus be assigned to the ability of the OREE to enhance the damping-like component of the SOT. While OREE is typically associated with a dominant FL torque, our switching experiments demonstrate its effectiveness as a powerful tool for probing the DLtorque and, more broadly, for torque metrology. Table~\ref{tabletorques} summarizes the measured values and relative contributions of both torque efficiencies, $\xi_{FL}$ and $\xi_{DL}$ efficiencies for the two samples.

\begin{widetext}
\begin{center}
\begin{table}[h!]
\centering
 \begin{tabular}{|c|c|c|}
    \hline
    &   $\xi_{FL}$   &  $\xi_{DL}$  \\
    \hline
    Ta(5)|Pt(8)|Co(0.9)|Al(1.4)|Pt(3)    &   \begin{tabular}{ccc} 
    SHE  &  OREE    &   total   \\
    \hline
    0.025 &  0.075   &   0.10   \\
    25 \%&  75 \%   &   100 \%
\end{tabular}                           &  
\begin{tabular}{ccc} 
    SHE  &  OREE    &   total   \\
    \hline
    0.0375 &  0.03   &   0.0675    \\
    56 \% &  44 \%  &   100 \%
\end{tabular} \\
    \hline
    Ta(5)|Pt(8)|Co(0.9)|Al(3)|Pt(3)      &   
\begin{tabular}{ccc} 
    SHE  &  OREE    &   total   \\
    \hline
    0.03 &  0.21   &   0.24   \\
    13 \%&  87 \%   &   100 \%
\end{tabular}       &  
\begin{tabular}{ccc} 
    SHE  &  OREE    &   total   \\
    \hline
    0.045 &  0.07   &   0.115    \\
    39 \% &  61 \%  &   100 \%
    \end{tabular} \\
    \hline
    \end{tabular}
    \caption{Values and relative contributions of OREE and Pt SHE in both torque $\xi_{FL}$ and $\xi_{DL}$ efficiencies for the two samples. For instance, in Ta(5)|Pt(8)|Co(0.9)|Al(3)|Pt(3), SHE induces 0.03 field like efficiency and OREE 0.21. OREE thus induces 87 \% of the field like torque.}
    \label{tabletorques}
\end{table}
\end{center}
\end{widetext}

\section{Conclusions}

Through second-harmonic measurement techniques, we have investigated the spin-orbit torque (SOT) and orbital torque (OT) occurring Co|Al interfaces in Ta5|Pt8|Co(t$_{Co}$)|Al(t$_{Al}$)|Pt3 emerging from the Co-Al bonding and the orbital momentum locking (OML) at this specific interface. We have extended our investigations to Ta5|Pt8|Co(t$_{Co}$)|Pt($t_{Pt}$)|Al3|Pt3 systems with Pt atomic layer intercalation  in-between Co|Al in order to modify the interface properties. It results that interleaved Pt layers has for effect to decrease the torque efficiency owing to the loss of the OML properties. Density-functional theory calculations of the linear response of both orbital-Rashba and torque tensors revealed the orbital origin of both effects with a magnitude strongly dependent on the Co-Al direct bonding at the scale of single atomic planes. Semi-classical model of the transverse component of the angular momentum we have developed allows to quantify both SHE and OREE contributions to the torques in the different sample series. This emphases thus on the major role played by the OREE on the magnetization switching as observed experimentally on nanoscaled pillars.

\section*{Methods} 

\subsection{first-principle calculations and structural details}

Our numerical investigation is based on density functional theory (DFT) using a localized basis set as defined within SIESTA ~\cite{siesta_method}. The out-of-equilibrium transport calculations use a full \textit{ab initio} DFT Hamiltonian matrices obtained directly from \textsc{SIESTA}~\cite{siesta_method} conveniently exploiting  the atom-centered double-$\zeta$ plus polarization (DZP) basis sets allowing its further manipulation for its usage with the Kubo formalism. We use the energy cutoff for real-space mesh of 400~Ry. The self-consistent SOC is introduced via the full off-site approximation~\cite{siesta_on-site_soc} using fully-relativistic norm-conserving pseudopotentials~\cite{tm_pseudopotentials}. The system Hamiltonian and overlap matrices are obtained after performing a full self-consistent cycle and treated after within a post-processing routine utilizing SISL as an interface tool~\cite{zerothi_sisl}. The DFT simulations were carried out considering twenty four layers made out from twelve Co layers and twelve layers of Al. The slab of Co was obtained by considering the [001] direction of the hpc structure whereas Al was grown along the [111] direction of the fcc structure. We use the Perdew-Burke-Ernzerhof (PBE)~\cite{gga,pbe} exchange-correlation functional. Relativistic effects (although small in this case) were introduced via the on-site approximation~\cite{siesta_on-site_soc} using fully-relativistic norm-conserving pseudopotentials~\cite{tm_pseudopotentials}. The system's ground state properties are obtained after performing a full self-consistent cycle converged a $(15 \times 15 \times 1)$  $\vec{k}$-points sampling of the Brillouin zone using a 25 $\AA$ vacuum distance between periodic images to avoid spurious effects. We used the conjugate gradient algorithm to minimize the atomic forces below 0.01~eV/{\AA}. The orbital texture, orbital and spin accumulations were therefore calculated by a home-made procedure using the Hamiltonian and overlap matrices. In order to obtain the optimized structure, we considered the relative electronic energy of the system as function of the system's lattice constant. We found that the lattice constant that minimizes the energy is around $\sim$ 2.94 $\AA$ which also displays the largest value of the orbital accumulation as it is commented in the main and below this section.

\vspace{0.1in}

\subsection{Second-harmonic measurement methods for SOT \& OT}

In order to evaluate the two components of the SOT, we employed combined first and second harmonic Hall measurement technique on 5~$\mu$m wide Hall bars with an AC excitation of 727~Hz~\cite{krishnia2023large}. We used the 'field-sweeping' technique/procedure adapted to PMA samples. This corresponds to the application of an in-plane magnetic field increasing in amplitude able to rotate the magnetization in the film plane up to the saturation, either applied along the current-line ('DL geometry') or transverse to it along the voltage detection line ('FL geometry'). The $1^{st}$ harmonic signal was used to measure both PHE and AHE signals to extract the exact anisotropy; and the $2^{nd}$ harmonic signal was recorded for the SOT and OT measurements after getting rid of the (minor) thermal component when required. We compared the results from the two series of multilayers with Al and Cu light element interfaced with Co: Pt(8)|Co($t_{\rm Co}$)|Al(1.4)|Pt(3) and Pt(8)|Co($t_{\rm Co}$)|Cu(1.4)|Pt(3). Starting from negligible torques in fully symmetric Pt(8)|Co(0.9)|Pt(8) stack as expected, the asymmetric \textit{'control'} sample Pt(8)|Co(0.9)|Pt(3) gives $B_{\text{DL}}$ = 0.80~$\pm$ 0.05~mT and $B_{\text{FL}}$ 0.58 $\pm$ 0.15~mT for a current density $J_{\text{Pt}}$ in Pt = $10^{11}$~A/m$^2$ (later used as a reference). To obtain integrated torques over the whole ferromagnetic layer of thickness $t_{\rm Co}$ and hence to be able to make accurate comparison, we define a normalized quantity by multiplying $B_{DL,FL}$ with $t_{\rm Co}$ for a $10^{11}$~A/m$^2$ current density in Pt. This corresponds to the equivalent torque values for an applied electric field $\mathbf{E}=\rho_{Pt}\times J_{Pt}=2.5 \times 10^4$~V/m owing to the Pt resistivity $\rho_{Pt}=25~\mu \Omega.$~cm.

\vspace{0.1in}

\subsection{nanopillars fabrication and current induced magnetization switching experiment}

We fabricated nanopillars for magnetization switching experiments via electron beam lithography and ion beam etching. We proceeded to two lithography steps. The first step consisted in the definition of 500 nm wide Hall bars into the thin films. The second step consisted in patterning a 250~nm $\times$ 300~nm pillar of the whole stack in the middle of the bars crossing and etching the rest of the Hall bar down to the bottom Pt layer. This pillars are made of the whole metallic stack, hence acting as a magnetic element laid on a SOT track. 

\vspace{0.1in}

Regarding the measurement protocol, we first saturated the magnetization out-of-plane along the positive axis before applying an in plane magnetic field $\hat{H}_x$ allowing deterministic SOT driven switching. Electrical pulses of different amplitudes and width were applied in one arm of the crossing. The perpendicular arm was used to measure $R_{xy}$ with a second pulse of 100~$\mu$s width and 100~$\mu$A amplitude. This $R_{xy}$ measurement enables us to probe the magnetization state through $R_{xy}$ AHE. Following this protocol, for a given width of "writing" pulses, we applied pulses of increasing amplitudes from 0 to a few mA to observe positive switching. We operated the reverse process with negative current to observe negative switching and complete magnetization switching cycles (figure \ref{fig:Switch}). We have extracted the critical switching current $I_c$ as the smallest current giving at least 70 \% magnetization reversal.

\begin{acknowledgments}
We thank S. Mallick for his help to prepare the nanopatterned samples. We are grateful to K. Garello (CEA-SPINTEC) for very fruitful discussions. N. Sebe benefits from a France 2030 government grant managed by the French National Research Agency (ANR-22-PEPR-0009 Electronique-EMCOM).  This study has been supported by the French National Research Agency under France 2030 government grant managed by the French National Research Agency PEPR SPIN ANR-22-EXSP0009 (SPINTHEORY) and by the EIC Pathfinder OPEN grant 101129641 ’OBELIX’. 
\end{acknowledgments}

\vspace{0.1in}

\section*{Data availability}

The data that support the findings of this study are available from the
corresponding author upon reasonable request.

\vspace{0.1in}

\section*{Author contributions}
N.S. prepare and design the samples, measured and provides data analyses, modeling and took part to the writing of the article. A.~P. provided DFT output, data analyses and took part to the writing. S. Krishnia measured a part of the sample series and provide data analyses. S. Collin prepared the samples. J.-M. took part to the analyses. A.-F. took part to the discussions and analyses. V.~C. and H.~J. managed the project, took part to the analyses and modeling and took part to the writing. 

\vspace{0.2in}

\section*{competing interest}
The authors declare no competing interests.

%\bibliography{biblioSebe.bib}
\bibliography{biblio.bib}

\end{document}